\documentclass{elsart}
\usepackage{graphicx,amssymb}

\begin{document}

\begin{frontmatter}


\title{Unusual field-induced transitions in exactly solved mixed spin-($1/2$, $1$) Ising chain with axial and rhombic zero-field splitting parameters\thanksref{grant}}       

\thanks[grant]{This work was supported by the ERDF EU (European Union European regional development fond) grant, under the contract No. ITMS26220120005 (activity 3.2.). This work was also supported  under the grant No.~VVGS~1/10-11.}
\author{Jozef Stre\v{c}ka},
\ead{jozef.strecka@upjs.sk}
\ead[url]{http://158.197.33.91/$\thicksim$strecka}
\author{Michal Dan\v{c}o} 
\address{Department of Theoretical Physics and Astrophysics, 
Faculty of Science, \\ P. J. \v{S}af\'{a}rik University, Park Angelinum 9,
040 01 Ko\v{s}ice, Slovak Republic}

\begin{abstract}
The mixed spin-(1/2, 1) Ising chain with axial and rhombic zero-field splitting parameters in a presence of the longitudinal magnetic field is exactly solved within the framework of decoration-iteration transformation and transfer-matrix method. Our particular emphasis is laid 
on an investigation of the influence of the rhombic term, which is responsible for an onset of quantum entanglement between two magnetic states $S_k^z = \pm 1$ of the spin-1 atoms. It is shown that 
the rhombic term gradually destroys a classical ferrimagnetic order in the ground state and simultaneously causes diversity in magnetization curves including intermediate plateau regions,  regions with a continuous change in the magnetization as well as several unusual field-induced transitions accompanied with magnetization jumps. Another interesting findings concern with an appearance of the round minimum in the temperature dependence of susceptibility times temperature data, the double-peak zero-field specific heat curves and the enhanced magnetocaloric effect. The temperature dependence of the specific heat with three separate maxima may also be detected when driving the system through the axial and rhombic zero-field splitting parameters close enough to a phase boundary between the ferrimagnetic and disordered states and applying sufficiently small longitudinal magnetic field. 
\end{abstract}
\end{frontmatter}
\journal{Physica B}

\section{Introduction}
\label{intro}

Over the last few decades, exactly solved one-dimensional quantum spin models \cite{baxt82,matt93,lieb04} have attracted considerable research interest as they may describe subtle quantum phenomena to emerge in real magnetic materials without a danger of over-interpretation, which is inherent to any approximative treatment. The present work is devoted to an exact study of the mixed spin-(1/2, 1) Ising chain model, which accounts both for the axial zero-field splitting (AZFS) as well as the rhombic zero-field splitting (RZFS) parameter in a presence of the applied longitudinal magnetic field. It is worthy of notice that the special limiting case of this model system in an absence of the external magnetic field has been proposed and exactly solved by Wu \textit{et al}. \cite{hain08,hain10,hain11} using the rigorous procedure based on the Jordan-Wigner transformation \cite{jord28} (see Refs. \cite{hain09,yang08,yang09} for related works on this subject). However, it has been recently shown by the present authors \cite{danc10} that the exact results obtained by Wu \textit{et al}. \cite{hain08,hain10} can also be recovered by another independent way by making use of the transfer-matrix method. The foremost advantage of the formulation based on the transfer-matrix method lies in the fact that this rigorous method may be even applied in a presence of the non-zero longitudinal magnetic field. The main purpose of this work is therefore to investigate the effect of longitudinal field on magnetic properties of the mixed spin-(1/2, 1) Ising chain with both AZFS and RZFS parameters. 

Before proceeding to an exact calculation for the investigated model system, let us briefly comment 
on an experimental motivation of our study. It is noteworthy that there exist several heterometallic  molecular-based compounds with a magnetic structure, which can be properly described as one-dimensional chain of alternating spin-$1/2$ and spin-$1$ metal ions. Among the most common examples of the one-dimensional mixed spin-(1/2, 1) chains one could mention
\begin{itemize}
    \item CuNi(EDTA).$6$H$_{2}$O, \cite{coro90}
    \item CuNi(pbaOH)(H$_{2}$O)$_{3}$.nH$_{2}$O, \cite{kahn90}
    \item CuNi(pba)(D$_{2}$O)$_{3}$.$2$D$_{2}$O, \cite{hagi98}
    \item PPh$_{4}$[Ni(pn)$_{2}$][Fe(CN)$_{6}$]H$_{2}$O, \cite{ohba98}
    \item \{Pr(bet)$_{2}$(H$_{2}$O)$_{3}$Fe(CN)$_{6}$\}, \cite{wang00}
    \item \{Ru(acac)$_{2}$(CN)$_{2}$\}\{Ni(dmphen)(NO$_{3}$)\}. \cite{toma06}       
\end{itemize}
Even though the vast majority of aforementioned polymeric compounds should be preferentially regarded as experimental representatives of the mixed-spin quantum Heisenberg chain rather than the mixed-spin Ising chain, it is the authors' hope that our exact analytical results for the mixed spin-(1/2, 1) Ising chain may provide a deeper insight into several important vestiges of real physical behavior and explain some experimental results at least at a qualitative level. Besides, one should also expect
that the theoretical description based on the mixed spin-$(1/2, 1)$ Ising chain may be quite appropriate for those heterometallic coordination polymers, where at least one from both constituent magnetic ions possesses a rather high magnetic anisotropy. It should be stressed that the magnetic behavior of this type has been recently found in two heterometallic complexes containing highly anisotropic rare-earth ions \cite{visi09,rinc10,heuv10}.

The outline of the present paper is as follows. In the following Section \ref{model}, we will shed light on the basic steps of our exact calculation for the investigated model system. The Section \ref{result} deals with the discussion of the most interesting results obtained for the phase diagrams and basic thermodynamic quantities. Finally, some concluding remarks are given in the Section \ref{conclusion}. 

\section{Exact solution of the mixed-spin Ising chain}
\label{model}

Let us consider the mixed spin-$(1/2, 1)$ Ising chain with AZFS and RZFS parameters in a presence of the longitudinal external magnetic field. Suppose that the linear chain consist of the alternating spin-$1/2$ and spin-$1$ atoms, whereas the former spin-$1/2$ atoms constitute the sublattice $A$ and the latter spin-$1$ atoms form the sublattice $B$. The total Hamiltonian of the system can be written as a sum of three parts
\begin{eqnarray}
\hat{\cal H} = \hat{\cal H}_{ex} + \hat{\cal H}_{zfs}^{(1)} + \hat{\cal H}_{zee},
\label{1}
\end{eqnarray}
which account for the nearest-neighbor Ising interaction, AZFS and RZFS terms acting on the spin-$1$ atoms and the magnetostatic (Zeeman's) energy of the spin-$1/2$ and spin-$1$ atoms in the applied longitudinal magnetic field
\begin{eqnarray}
\hat{\cal H}_{ex} &=& -J \sum_{k=1}^{N} \hat{S}_{k}^z (\hat{\sigma}_{k}^z + \hat{\sigma}_{k+1}^z), \label{2a} \\
\hat{\cal H}_{zfs}^{(1)} &=& -D \sum_{k=1}^{N} (\hat{S}_{k}^z)^2 - E \sum_{k=1}^{N} [(\hat{S}_{k}^x)^2 - (\hat{S}_{k}^y)^2], \label{2b} \\
\hat{\cal H}_{zee} &=& -H_{A} \sum_{k=1}^{N} \hat{\sigma}_{k}^z - H_{B} \sum_{k=1}^{N} \hat{S}_{k}^z.
\label{2c}
\end{eqnarray}
Above, $\hat{\sigma}_{k}^z$ and $\hat{S}_{k}^\alpha ~(\alpha=x,y,z)$ denote standard spatial components of the spin-1/2 and spin-1 operators, respectively, $N$ denotes a total number of spins from each sublattice and the periodic boundary condition $\sigma_{N+1} \equiv \sigma_{1}$ is imposed for simplicity. The parameter $J$ stands for the Ising interaction between nearest-neighboring spin-$1/2$ and spin-$1$ atoms, whereas the terms $D$ and $E$ label the AZFS and RZFS parameters acting on the spin-$1$ atoms only. Last, two Zeeman's terms $H_{A}$ and $H_{B}$ describe the influence of longitudinal magnetic field on the spin-$1/2$ and spin-$1$ atoms from the sublattice $A$ and $B$, respectively.

Before proceeding further, it is worthwhile to remark that there exist several equivalent representations of the zero-field splitting Hamiltonian $\hat{\cal H}_{zfs}^{(1)}$ given by Eq. (\ref{2b}). As a matter of fact, one may for instance prove one-to-one correspondence between $\hat{\cal H}_{zfs}^{(1)}$ and the effective spin Hamiltonian with three different single-ion anisotropy parameters $D^x$, $D^y$ and $D^z$
\begin{eqnarray}
\hat{\cal H}_{zfs}^{(2)} = -D^x \sum_{k=1}^{N} (\hat{S}_{k}^x)^2 - D^y \sum_{k=1}^{N} (\hat{S}_{k}^y)^2 - D^z \sum_{k=1}^{N} (\hat{S}_{k}^z)^2.
\label{3}
\end{eqnarray}
The Hamiltonians $\hat{\cal H}_{zfs}^{(1)}$ and $\hat{\cal H}_{zfs}^{(2)}$ differ one from the other just by unimportant constant term, because the mapping relations $D = D^z - \frac{D^x + D^y}{2}$ and $E = \frac{D^x - D^y}{2}$ establish a precise equivalence between the Hamiltonians (\ref{2b}) and (\ref{3}) (see Ref. \cite{danc10} for more details). It should be also noticed that the particular case of the Hamiltonian $\hat{\cal H}_{zfs}^{(2)}$ with $D^y=0$ has been considered by Wu \textit{et al.} \cite{hain08} in their recent work. However, it has been shown in our preliminary report \cite{danc10} that the Hamiltonian $\hat{\cal H}_{zfs}^{(1)}$ with one less free parameter is much more appropriate for the interpretation of obtained exact results compared with the Hamiltonian $\hat{\cal H}_{zfs}^{(2)}$ and thus, this more convenient definition of the zero-field-splitting Hamiltonian will be used throughout the rest of this paper.

Now, let us turn our attention to the main points of the method, which enables an exact treatment of the investigated quantum spin chain. First, the total Hamiltonian (\ref{1}) can be rewritten as a sum of the Zeeman's term for all spin-1/2 atoms from the sublattice $A$ and the sum of site Hamiltonians
\begin{eqnarray}
\hat{\cal H} = \sum_{k=1}^{N} \hat{\cal H}_{k} - H_{A}\sum_{k=1}^{N} \hat{\sigma}_{k}^z,
\label{4}
\end{eqnarray}
whereas each site Hamiltonian $\hat{\cal H}_{k}$ involves all the interaction terms including the $k$th spin-$1$ atom from the sublattice $B$
\begin{eqnarray}
\hat{\cal H}_{k} = -E_{k} \hat{S}_{k}^z - D(\hat{S}_{k}^z)^2 - E[(\hat{S}_{k}^x)^2 - (\hat{S}_{k}^y)^2]
\label{5}
\end{eqnarray}
with $E_{k} = J(\hat{\sigma}_{k}^z + \hat{\sigma}_{k+1}^z) + H_{B}$. Because the Hamiltonians $(7)$ at different sites commute, i.e. $[\hat{\cal H}_{i},\hat{\cal H}_{j}]=0$ is valid for each $i\neq j$, the partition function can be partially factorized and consequently rewritten into the form
\begin{eqnarray}
{\cal Z} = \sum_{\{\sigma_{k}\}}\prod_{k=1}^{N} \exp \left[\frac{\beta H_{A}}{2}(\hat{\sigma}_{k}^z + \hat{\sigma}_{k+1}^z)\right]\mbox{Tr}_{S_{k}}\exp (-\beta \hat{\cal H}_{k}) = \sum_{\{\sigma_{k}\}}\prod_{k=1}^{N} {\cal Z}_k,
\label{6}
\end{eqnarray}
where $\beta = 1/(k_{\rm B} T)$, $k_{\rm B}$ is Boltzmann's constant, $T$ is the absolute temperature, $\mbox{Tr}_{S_{k}}$ means a trace over degrees of freedom of the $k$th spin-1 atom from the sublattice $B$ and the symbol $\sum_{\{\sigma_{k}\}}$ denotes a summation over all possible configurations of the spin-$1/2$ atoms from the sublattice $A$. The crucial step of our exact procedure represents calculation of the expression $\mbox{Tr}_{S_{k}}\exp (-\beta \hat{\cal H}_{k})$. For this purpose, it is useful to rewrite the site Hamiltonian (\ref{5}) into the usual matrix representation
\begin{eqnarray}
\left\langle k_{i}\right| \hat{\cal H}_{k} \left|k_{j}\right\rangle = 
\left( \begin{array}{ccc}
-E_{k}-D & 0 & -E \\
     0   & 0 &  0 \\
-E       & 0 & E_{k}-D
\end{array} \right)   
\label{7}
\end{eqnarray}
using the standard basis of ket vectors $\left|k_{j}\right\rangle = \left|\pm1\right\rangle,\left|0\right\rangle$ ($j=1-3$) corresponding to the three possible spin states $S_{k}^z=\pm1,0$ of the $k$th spin-$1$ atom from the sublattice $B$. The straightforward diagonalization of the site Hamiltonian yields the following eigenenergies and eigenvectors
\begin{eqnarray}
\lambda_{k1} &=& -D-\sqrt{E_{k}^2 + E^2}, \qquad 
\left|\psi_{k1}\right\rangle = \cos \left(\frac{\varphi_k}{2}\right) \left|1\right\rangle 
                            + \sin \left(\frac{\varphi_k}{2}\right) \left|-1\right\rangle, \nonumber \\ 
\lambda_{k2} &=&  0, \,\;\,\;\qquad \qquad \qquad \qquad 
\left|\psi_{k2}\right\rangle = \left|0\right\rangle,  \nonumber \\
\lambda_{k3} &=& -D+\sqrt{E_{k}^2 + E^2}, \qquad 
\left|\psi_{k3}\right\rangle = \sin \left(\frac{\varphi_k}{2}\right) \left|1\right\rangle 
                            + \cos \left(\frac{\varphi_k}{2}\right) \left|-1\right\rangle, 
\label{8}
\end{eqnarray}
with the mixing angle $\varphi_k$ defined through the relation $\varphi_k = \arctan (\frac{E}{E_{k}})$. It is worth mentioning that the eigenenergies listed in the set of Eqs. (\ref{8}) can readily be used for calculating the expression $\mbox{Tr}_{S_{k}}\exp (-\beta \hat{\cal H}_{k})$ and moreover, the analytical form of the site partition function ${\cal Z}_{k}$ then immediately implies a possibility of performing the generalized decoration-iteration transformation \cite{fish59,roja09,stre10}
\begin{eqnarray}
{\cal Z}_{k} &=& \exp \left[\frac{\beta H_{A}}{2} \left(\sigma_{k}^z + \sigma_{k+1}^z \right) \right]
    \left[1 + 2\exp(\beta D)\cosh\left(\beta \sqrt{E_{k}^2 + E^2}\right) \right] \nonumber \\
&=& A\exp\left[\beta R \sigma_{k}^z \sigma_{k+1}^z + \frac{\beta H_{0}}{2}(\sigma_{k}^z + \sigma_{k+1}^z)\right].
\label{9}
\end{eqnarray}
The unknown mapping parameters $A$, $R$ and $H_{0}$ entering in Eq. (\ref{9}) can be obtained from the self-consistency condition of the decoration-iteration transformation. Following the standard procedure \cite{fish59,roja09,stre10}, one directly obtains explicit expressions for all three transformation parameters
\begin{eqnarray}
A = (V_{1}V_{2}V_{3}^2)^\frac{1}{4} , \qquad \beta H_{0} = \ln\left(\frac{V_{1}}{V_{2}}\right) , \qquad \beta R = \ln\left( \frac{V_{1}V_{2}}{V_{3}^2}\right),
\label{10}
\end{eqnarray}
which are defined in terms of newly defined functions $V_{1}$, $V_{2}$ and $V_{3}$
\begin{eqnarray}
V_{1} &=& \exp \left(\frac{\beta H_{A}}{2}\right)\left[1 + 2\exp(\beta D)\cosh\left(\beta \sqrt{(J + H_{B})^2 + E^2}\right)\right],  \nonumber \\
V_{2} &=& \exp \left(-\frac{\beta H_{A}}{2}\right)\left[1 + 2\exp(\beta D)\cosh\left(\beta \sqrt{(J - H_{B})^2 + E^2}\right)\right],  \nonumber  \\
V_{3} &=& 1 + 2\exp(\beta D)\cosh\left(\beta \sqrt{H_{B}^2 + E^2}\right).
\label{11}
\end{eqnarray}
The functions $V_1$, $V_2$ and $V_3$ physically correspond to three independent expressions for the site partition function ${\cal Z}_{k}$, which can be obtained from the transformation formula (\ref{9}) by considering all four spin configurations of two Ising spins $\sigma_{k}$ and $\sigma_{k+1}$ involved therein (recall that they enter into the parameter $E_k$ as well). The substitution of the decoration-iteration transformation (\ref{9}) into the formula (\ref{6}) leads in turn to a rigorous mapping relationship between the partition function ${\cal Z}$ of the mixed-spin Ising chain and the partition function ${\cal Z}_{ic}$ of the simple spin-$1/2$ Ising chain with the effective temperature-dependent nearest-neighbor interaction $R$ and magnetic field $H_0$
\begin{eqnarray}
{\cal Z}(\beta, J, D, E, H_{A}, H_{B}) = A^N {\cal Z}_{ic}(\beta, R, H_{0}).
\label{12}
\end{eqnarray}
The relationship (\ref{12}) between both partition functions proves that there exist a rigorous mapping correspondence between the mixed-spin Ising chain and the simple spin-$1/2$ Ising linear chain, since both partition functions differ one from another just by the multiplicative factor $A$ given by Eqs. (\ref{10})-(\ref{11}). In this regard, the equality (\ref{12}) may be considered as a mathematical proof of the aforementioned mapping equivalence and hence, the partition function of the mixed-spin Ising chain can exactly be calculated from the well-known formula for the partition function of the simple spin-1/2 Ising linear chain \cite{baxt82} with the effective nearest-neighbor interaction $R$ and magnetic field $H_0$ unambiguously determined by Eqs. (\ref{10})-(\ref{11}). Besides, the mapping relationship (\ref{12}) is also very suitable for deriving exact results for other quantities, such as free energy, magnetization, entropy, specific heat, susceptibility, etc. For instance, the free energy of the mixed-spin Ising chain per unit cell reads
\begin{eqnarray}
f = \frac{F}{N} =  k_{\rm B}T\ln2 - k_{\rm B}T\ln\left(V_{1} + V_{2} + \sqrt{(V_{1} - V_{2})^2 + 4V_{3}^2}\right)
\label{13}
\end{eqnarray}
At this stage, both sublattice magnetizations can easily be calculated as a partial derivative of the the free energy with respect to the relevant magnetic field, namely, $m_{A}=- \left(\frac{\partial f}{\partial H_{A}} \right)_T$ and $m_{B}=- \left(\frac{\partial f}{\partial H_{B}} \right)_T$. 
After straightforward, but a little bit cumbersome arrangement, both sublattice magnetizations 
can be written in this compact form
\begin{eqnarray}
m_{A} &=& \left\langle \hat{\sigma}_{k}^z\right\rangle = \frac{1}{2}\frac{V_{1} - V_{2}}{\sqrt{(V_{1} - V_{2})^2 + 4V_{3}^2}} \label{14a} \\
m_{B} &=& \left\langle \hat{S}_{k}^z\right\rangle \nonumber \\
&=& \frac{(V_{1}-V_{2})(W_{1}-W_{2}) + 4V_{3}W_{3} + (W_{1}+W_{2})\sqrt{(V_{1}-V_{2})^2 + 4V_{3}^2}}{(V_{1}-V_{2})^2 + 4V_{3}^2 + (V_{1}+V_{2})\sqrt{(V_{1}-V_{2})^2 + 4V_{3}^2}},  
\label{14b}
\end{eqnarray}
where $\langle \cdots \rangle$ means the standard canonical ensemble averaging and the functions $W_{1}$, $W_{2}$ and $W_{3}$ are defined as follows
\begin{eqnarray}
W_{1} &=& 2 \frac{\left(J + H_{B} \right)}{\sqrt{(J + H_{B})^2 + E^2}} \exp \left(\beta D+\frac{\beta H_{A}}{2}\right) \sinh\left[\beta \sqrt{(J + H_{B})^2 + E^2}\right]\!, \nonumber \\
W_{2} &=& - 2 \frac{\left(J - H_{B} \right)}{\sqrt{(J - H_{B})^2 + E^2}} \exp \left(\beta D-\frac{\beta H_{A}}{2}\right) \sinh\left[\beta \sqrt{(J - H_{B})^2 + E^2}\right]\!, \nonumber \\
W_{3} &=& 2 \frac{H_{B}}{\sqrt{H_{B}^2 + E^2}} \exp(\beta D) 
\sinh\left(\beta \sqrt{H_{B}^2 + E^2}\right)\!.
\label{15}
\end{eqnarray}
The total magnetization normalized with respect to its saturation value is then given by $m/m_s = 2(m_{A} + m_{B})/3$. Other basic thermodynamic quantities such as entropy and specific heat can 
also be readily calculated from the free energy (\ref{13}) with the help of basic thermodynamic relations
\begin{eqnarray}
S = -\left(\frac{\partial F}{\partial T}\right)_{H}, \qquad  
C = -T \left(\frac{\partial^2 F}{\partial T^2}\right)_{H}. 
\label{16}
\end{eqnarray} 
Notice that the final expressions for the entropy and specific heat obtained by the use of Eq. (\ref{16}) are too cumbersome to write them down here explicitly. Furthermore, one may also obtain 
the relevant exact result for the initial longitudinal susceptibility reduced per elementary unit 
from the previously derived exact results (\ref{14a})--(\ref{14b}) for both sublattice magnetizations   
\begin{eqnarray}
\chi_{z} &=& \left(\frac{\partial m_A}{\partial H_A}\right)_{T, H_A \rightarrow 0} 
         + \left(\frac{\partial m_B}{\partial H_B}\right)_{T, H_B \rightarrow 0} \nonumber \\
    &=& \beta \left[ \frac{U_1}{4 U_2} + \frac{U_3^2 + U_2 (U_4 + U_5 + U_6)}{U_2 (U_1 + U_2)} \right].
\label{is}
\end{eqnarray} 
For simplicity, the functions $U_k$ ($k = 1-6$) to emerge in Eq. (\ref{is}) for the initial longitudinal susceptibility are explicitly given in the Appendix. 

On the other hand, the exact results derived for the free energy (\ref{13}) or both sublattice magnetizations (\ref{14a})--(\ref{14b}) cannot be utilized for calculating the initial transverse susceptibility. For this purpose, one first needs to calculate the relevant spatial component of the quadrupolar moment in the limit of zero external magnetic field ($H_A = H_B = 0.0$) and to relate it subsequently to the initial transverse susceptibility using the fluctuation-dissipation theorem \cite{fish63}. It is noteworthy that all three spatial components of the quadrupolar moment can easily be obtained with the help of the generalized Callen-Suzuki identity \cite{call63,suzu65}. According to this, the statistical mean values determining the spatial components of the quadrupolar moment of the spin-1 atoms may be calculated from the exact spin identity
\begin{eqnarray}
q_{B}^\alpha = \left\langle (\hat{S}_{k}^\alpha)^2 \right\rangle = \left\langle \frac{\mbox{Tr}_{S_{k}}[(\hat{S}_{k}^\alpha)^2\exp(-\beta \hat{{\cal H}}_{k})]} {\mbox{Tr}_{S_{k}}\exp(-\beta \hat{{\cal H}}_{k})} \right\rangle. \qquad (\alpha = x, y, z)
\label{17}
\end{eqnarray}
After straightforward but rather lengthly calculation based on the exact spin identity (\ref{17}) and the relevant eigenvalues and eigenvectors (\ref{8}) of the site Hamiltonian (\ref{5}), all three spatial components of the quadrupolar moment can be expressed through the following formulas
\begin{eqnarray}
q_{B}^x = \frac{U_1^{+} + U_2^{+}}{U_1 + U_2}, \quad
q_{B}^y = \frac{U_1^{-} + U_2^{-}}{U_1 + U_2}, \quad
q_{B}^z = \frac{U_1 + U_2 - 2}{U_1 + U_2}, 
\label{19}
\end{eqnarray}
where the functions $U_i$ and $U_{i}^j$ are defined for $i=1,2$ and $j=\pm$ in Appendix. Now, both spatial components of the initial transverse susceptibility reduced per elementary unit can be easily calculated by making use of the fluctuation-dissipation theorem
\begin{eqnarray}
\chi_{\gamma} = \frac{\beta}{N} \left\{\left\langle \left[\sum_{k=1}^N(\hat{S}_{k}^\gamma + \hat{\sigma}_{k}^\gamma)\right]^2\right\rangle - \left\langle \left[\sum_{k=1}^N(\hat{S}_{k}^\gamma + \hat{\sigma}_{k}^\gamma)\right]\right\rangle^2\right\}. \quad (\gamma = x, y)
\label{20}
\end{eqnarray}
In the zero-field limit, the exact expression for the initial transverse susceptibility resulting 
from the fluctuation-dissipation theorem largely simplifies, because all nearest-neighbor as well as further-neighbor pair correlation functions vanish in an absence of pair interactions between transverse ($x$ or $y$) spatial components of the spin operators. Both spatial components of the initial transverse susceptibility can be thus expressed solely in terms of the respective spatial components of the quadrupolar moment
\begin{eqnarray}
\chi_{\gamma} = \beta\left[\frac{1}{4} +  \left\langle (S_{k}^\gamma)^2\right\rangle\right]=\beta\left(\frac{1}{4} + q_{B}^\gamma\right), \quad (\gamma = x, y)
\label{24}
\end{eqnarray}
which are exactly known from Eq. (\ref{19}) derived previously using the generalized 
Callen-Suzuki identity \cite{call63,suzu65}.

\section{Results and discussion}
\label{result}
In this section, we will focus our attention to the most interesting results to be obtained for phase diagrams and basic thermodynamic quantities. It is worthwhile to remark that the magnetic behavior of the investigated model system in an absence of the external magnetic field has been thoroughly examined in Refs. \cite{hain08,hain10,danc10} and thus, the respective behavior in a non-zero longitudinal magnetic field will be the main focus of the present paper. It should be also noticed that all results obtained in the preceding section are quite general as they hold both for the special case with the ferromagnetic ($J>0$) as well as antiferromagnetic ($J<0$) interaction. The ferromagnetic interaction between the nearest-neighbor spins generally leads to the ferromagnetic spin alignment, while the antiferromagnetic interaction causes the ferrimagnetic spin arrangement. It could be expected that the magnetic behavior of the ferrimagnetic model in the applied longitudinal magnetic field should be much more diverse compared with its ferromagnetic counterpart and therefore, we will henceforth assume 
$J<0$ for the sake of simplicity. In order to reduce the total number of free Hamiltonian parameters, our subsequent analysis will be restricted to the special case with the uniform magnetic field $H_{A}=H_{B}=H$ acting on the spin-$1/2$ and spin-$1$ atoms, which physically corresponds to the situation with the equal $g$-factors for both kinds of the magnetic atoms.

\subsection{Ground-state phase diagram}
Let us construct first the ground-state phase diagram, which will clarify the magnetic behavior of the ferrimagnetic mixed spin-(1/2,1) Ising chain with the AZFS and RZFS parameters in the longitudinal magnetic field.
\begin{figure}
\vspace{-0.8 cm}
\includegraphics[width=0.49\textwidth]{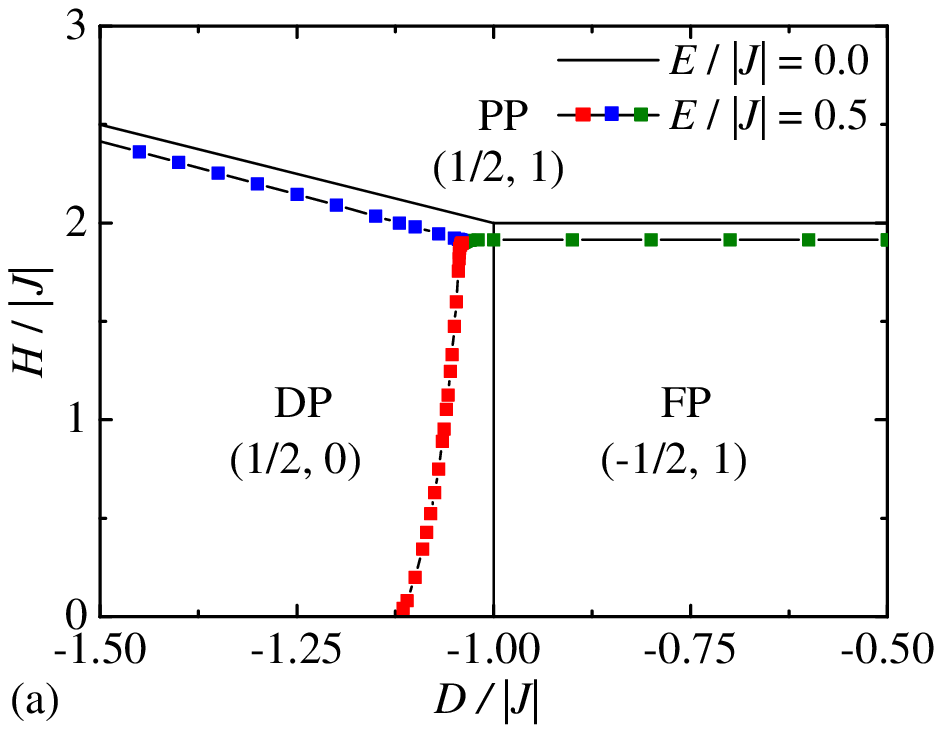}
\includegraphics[width=0.49\textwidth]{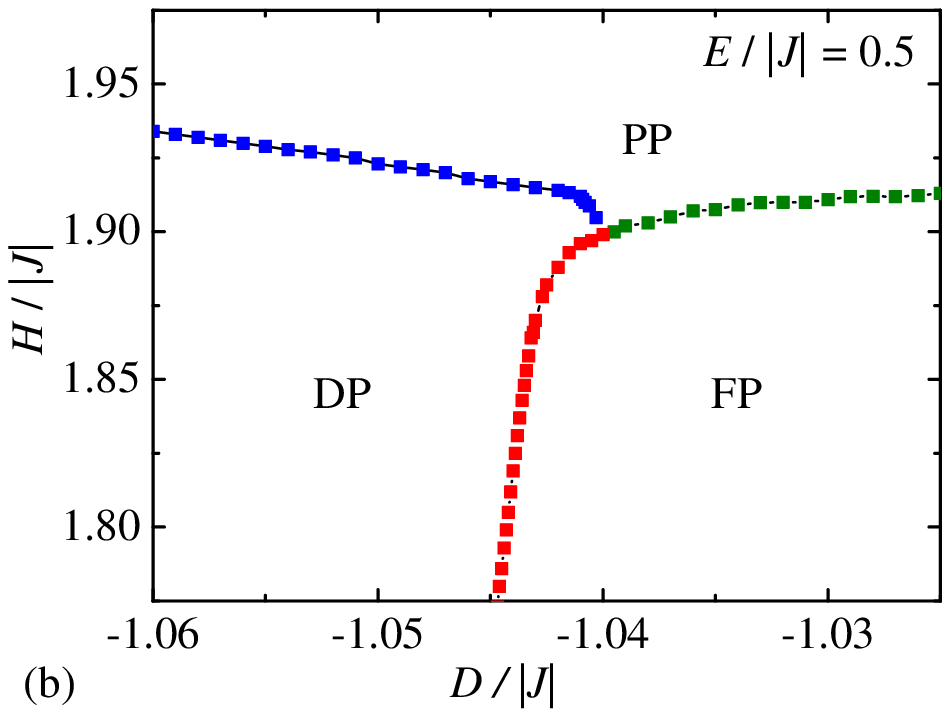}
\vspace{-0.7cm}
\caption{The ground-state phase diagram in the $D/\left|J\right| - H/\left|J\right|$ plane of the ferrimagnetic ($J<0$) mixed spin-(1/2,1) Ising chain for: (a) two different values of RZFS parameter $E/\left|J\right|=0.0$ (solid lines) and $E/\left|J\right|=0.5$ (solid lines with symbols), (b) the detail 
of the particular case with $E/\left|J\right|=0.5$ in the vicinity of the triple point.}  
\label{FD}
\end{figure}
The ground-state phase diagram in the $D/\left|J\right| - H/\left|J\right|$ plane is illustrated in Fig. \ref{FD}(a) for two different values of the RZFS parameter $E/\left|J\right|=0.0$ and $0.5$. 
As one can see, the investigated mixed-spin chain keeps under the influence of longitudinal magnetic field 
one from three available ground states, which are separated one from another by discontinuous (first-order) phase transition lines. Three spin arrangements emerging within the available 
ground states can be described through the following eigenvectors. 
\newline
Quantum ferrimagnetic phase:
\begin{eqnarray}
\left| FP \right\rangle &=& \prod_{k=1}^N\left|\sigma_{k}^z= -\frac{1}{2} \right\rangle \left[\cos\left(\frac{\varphi_{1}}{2}\right)\left|S_{k}^z=1\right\rangle + \sin\left(\frac{\varphi_{1}}{2}\right)\left|S_{k}^z=-1\right\rangle\right], \nonumber \\
\varphi_{1} &=& \arctan\left(\frac{E}{H_{B}+\left|J\right|}\right), 
\label{26}
\end{eqnarray}
Disordered phase:
\begin{eqnarray}
\left| DP \right\rangle &=& \prod_{k=1}^N \left| \sigma_{k}^z= {\rm sgn}(H)\frac{1}{2} \right\rangle \left| S_{k}^z=0 \right\rangle,
\label{27}
\end{eqnarray}  
Quantum paramagnetic phase:
\begin{eqnarray}
\left| PP \right\rangle &=& \prod_{k=1}^N\left| \sigma_{k}^z= \frac{1}{2} \right\rangle \left[\cos\left(\frac{\varphi_{2}}{2}\right)\left|S_{k}^z=1\right\rangle + \sin\left(\frac{\varphi_{2}}{2}\right)\left|S_{k}^z=-1\right\rangle\right],  \nonumber \\
\varphi_{2} &=& \arctan\left(\frac{E}{H_{B}-\left|J\right|}\right). 
\label{25}
\end{eqnarray}

The spin alignment becomes quite obvious when assuming the particular case without the RZFS anisotropy, 
i.e. $E/\left|J\right|=0.0$. The eigenvector $\left| FP \right\rangle$ then refers to the classical ferrimagnetic phase with a perfect antiparallel orientation of the nearest-neighboring spin-1/2 and spin-1 atoms each of them occupying the spin state $\left|\sigma_{k}^z=-1/2\right\rangle$ and $\left|S_{k}^z=1\right\rangle$, respectively. 
This phase appears in the ground state just if the AZFS parameter is greater than the boundary value $D_{b}/\left|J\right|= -1.0$, otherwise the disordered phase $\left| DP \right\rangle$ becomes the ground state. In an absence of the external magnetic field, each spin-1/2 atom randomly occupies 
in $\left| DP \right\rangle$ any of its two available spin states $\left|\sigma_{k}^z=\pm 1/2\right\rangle$ on behalf of the non-magnetic character $\left|S_{k}^z=0\right\rangle$ of all the spin-$1$ atoms, which is enforced by the AZFS parameter. In any non-zero field, all the spins-$1/2$ atoms tend to align into the external-field direction as they effectively behave as free paramagnetic spins within $\left| DP \right\rangle$. Finally, the classical paramagnetic phase $\left| PP \right\rangle$ becomes the ground state whenever the external field exceeds the saturation value $H_{s}/\left|J\right|= 2.0$ for $D/\left|J\right|> -1.0$ or $H_{s}/\left|J\right|= 1 - D/|J|$ 
for $D/\left|J\right|< -1.0$. The classical paramagnetic phase can be characterized by a perfect alignment of all the spin-1/2 as well as spin-1 atoms into the external-field direction. 
All three possible ground states coexist together at a triple point given by the coordinates $[D_t/\left|J\right|, H_t/\left|J\right|]=[-1.0, 2.0]$.

Contrary to this, the situation becomes much more involved when assuming the non-zero RZFS parameter and its influence on the ground-state spin arrangement. The typical ground-state phase diagram is shown in Fig. \ref{FD}(a) for one specific value of the RZFS parameter $E/\left|J\right|=0.5$. The non-zero RZFS term introduces into the otherwise classical mixed-spin Ising chain $x$- and $y$-components of the spin operators and hence, this parameter is responsible for the onset of local quantum fluctuations that consequently lead to a quantum entanglement between two magnetic spin states $\left|S_{k}^z= \pm1\right\rangle$ of the spin-1 atoms. In this respect, the RZFS parameter gradually destroys the perfect ferrimagnetic spin arrangement, which exists in $\left| FP \right\rangle$ in an absence of the RZFS term. Indeed, the quantum superposition of the spin states $\left|S_{k}^z=1\right\rangle$ and $\left|S_{k}^z=-1\right\rangle$ governs the overall magnetic behavior of the spin-$1$ atoms 
in $\left| FP \right\rangle$, whereas the relevant spin state $\left|\sigma_{k}^z=- 1/2\right\rangle$ of the spin-1/2 atoms is not affected by the RZFS term at all. It is also quite evident from the mixing angle $\varphi_{1}$ given by Eq.~(\ref{26}) that the RZFS term generally enhances the probability amplitude of the less probable spin state $\left|S_{k}^z=-1\right\rangle$ on account of the more probable spin state $\left|S_{k}^z=1\right\rangle$, while the reverse is the case when considering the effect of the longitudinal magnetic field. If the external field exceeds some critical value, the investigated spin system passes from the quantum ferrimagnetic phase $\left| FP \right\rangle$ to the quantum paramagnetic phase $\left| PP \right\rangle$. All the spin-$1/2$ atoms tend to align towards the external-field direction in $\left| PP \right\rangle$, while the quantum entanglement of the spin states $\left|S_{k}^z=1\right\rangle$ and $\left|S_{k}^z=-1\right\rangle$ persists in this phase even if the rising magnetic field gradually suppresses the quantum superposition defined through 
the another mixing angle $\varphi_{2}$ given by Eq.~(\ref{25}). 

As far as the disordered phase $\left| DP \right\rangle$ is concerned, there is no change in the 
relevant spin arrangement of this phase in comparison with the previously discussed specific case 
without the RZFS term. Namely, the RZFS parameter does not affect the overall spin arrangement 
inherent to $\left| DP \right\rangle$, since this term is responsible only for the quantum entanglement between two magnetic spin states $\left|S_{k}^z=\pm1\right\rangle$, whereas all the spin-$1$ atoms reside the non-magnetic spin state $\left|S_{k}^z=0\right\rangle$ in $\left| DP \right\rangle$. However, there is an interesting shift of the ground-state phase boundary between $\left| FP \right\rangle$ and $\left| DP \right\rangle$ to a more negative values of the AZFS parameter, which emerges as a result of the influence of the RZFS parameter. In fact, this phase boundary becomes a striking non-linear curve instead of a simple vertical line for any non-zero RZFS parameter [see the detail depicted in Fig.~\ref{FD}(b) as well]. In a relatively narrow range of the AZFS parameters, e.g. $D/\left|J\right|\in(-1.115,-1.040)$ for the particular case with $E/\left|J\right|=0.5$, the investigated mixed-spin chain thus displays a peculiar sequence of two field-induced phase transitions. At lower critical field, the system first undergoes the field-induced transition from $\left| FP \right\rangle$ to $\left| DP \right\rangle$, while another field-induced transition from $\left| DP \right\rangle$ to $\left| PP \right\rangle$ takes place at upper critical field. The triple point, at which all three phases coexist together in the ground state, is given by the coordinates [$D_{t}/\left|J\right|,H_{t}/\left|J\right|$] = [$-1.040$,$1.899$] for the specific case with $E/\left|J\right|=0.5$. 

\subsection{Magnetization process}

Now, let us explore in detail the effect of AZFS and RZFS parameters on the magnetization process. 
\begin{figure} 
\vspace{-0.8 cm}
\begin{center}
\includegraphics[width = 0.49\textwidth]{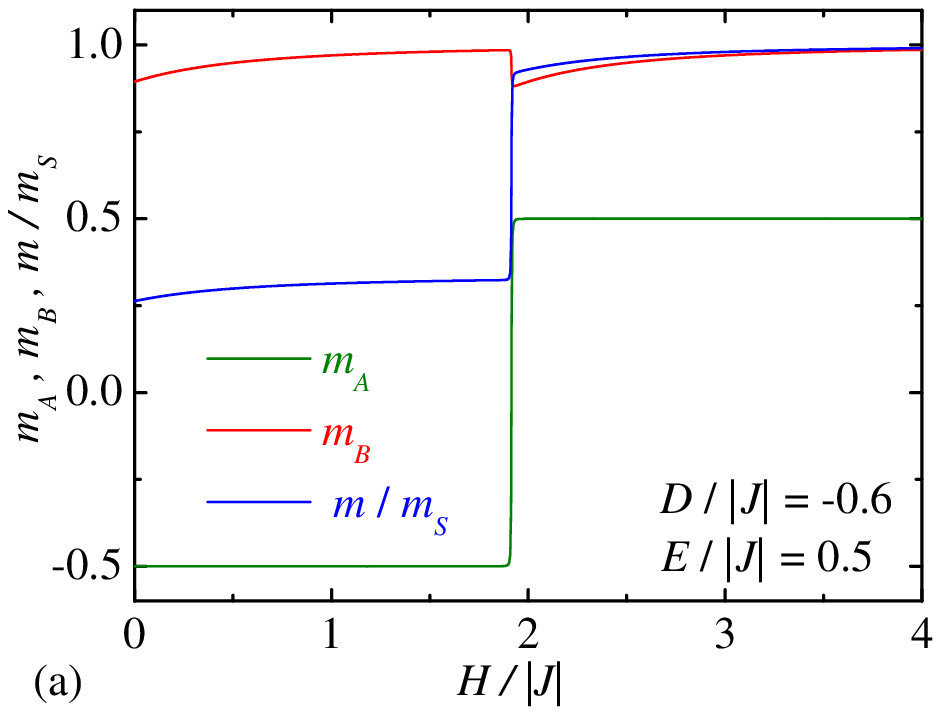}
\includegraphics[width = 0.49\textwidth]{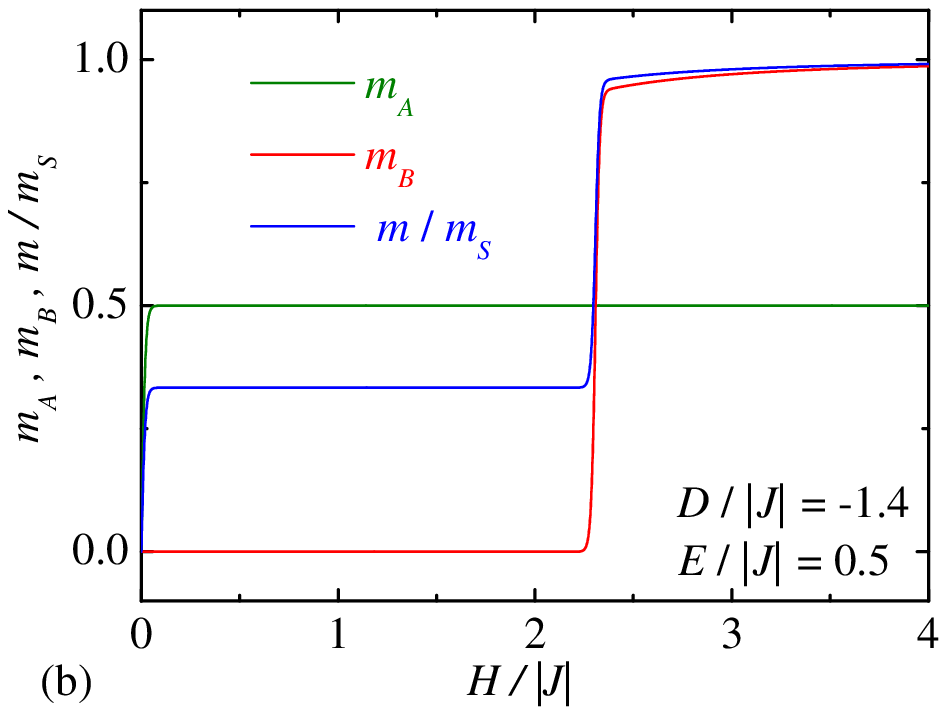}
\includegraphics[width = 0.49\textwidth]{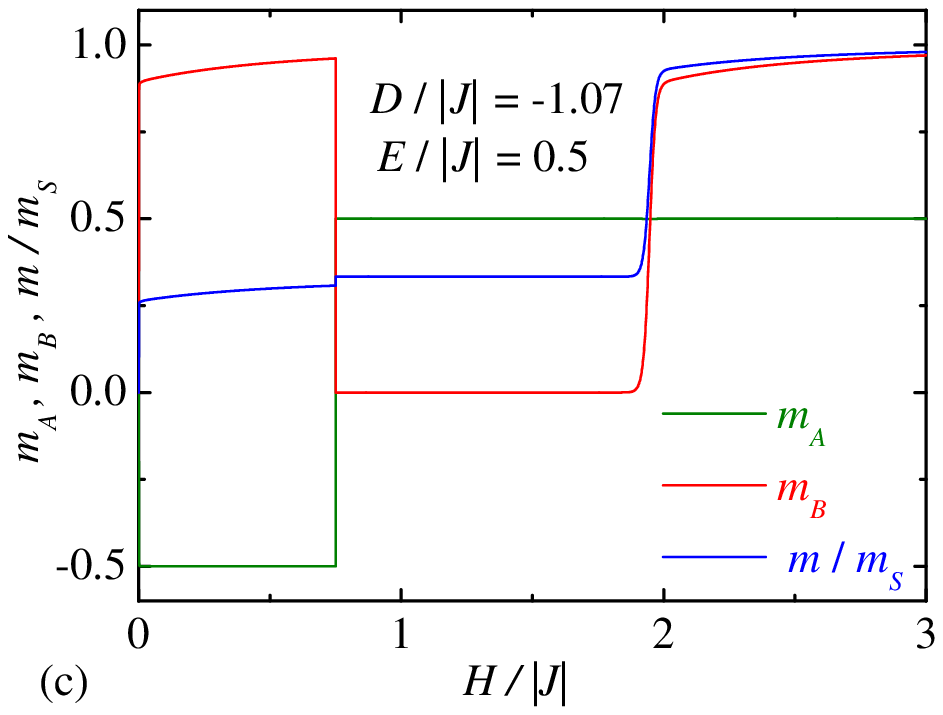}
\includegraphics[width = 0.49\textwidth]{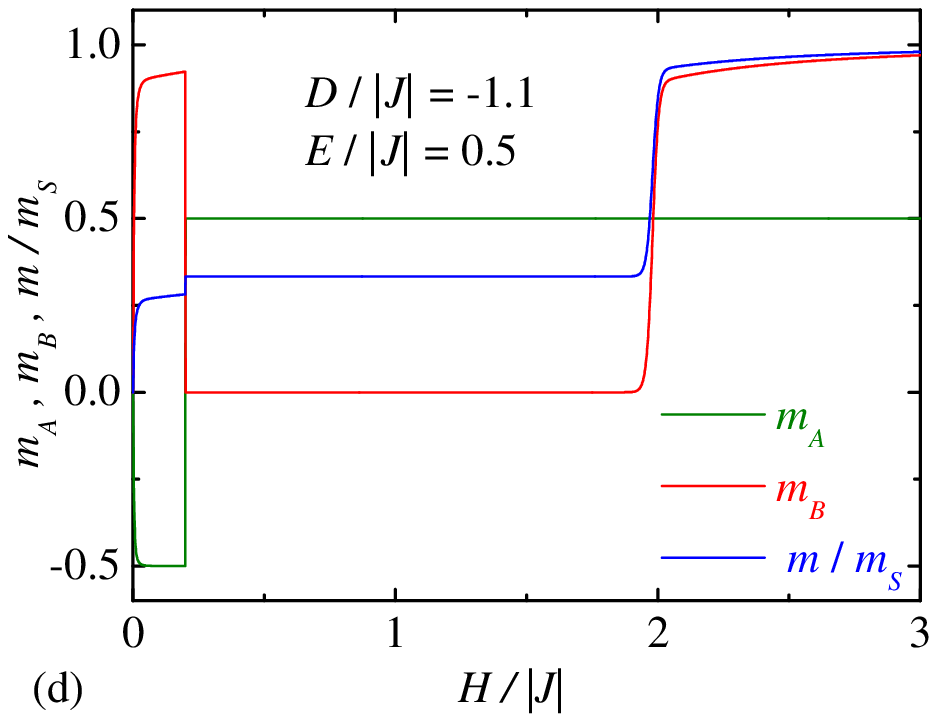}
\vspace{-0.6 cm}
\caption{\small Both sublattice magnetizations as well as the total magnetization reduced with respect to its saturation value versus the external magnetic field at the sufficiently low temperature $k_{B}T/\left|J\right|=0.01$, the fixed value of the RZFS parameter $E/\left|J\right|=0.5$ and 
four different values of the AZFS parameter $D/\left|J\right|= -0.6,-1.4,-1.07, -1.1$.}
\label{MAG}
\end{center}
\end{figure}
Both sublattice magnetizations ($m_{A}$, $m_{B}$) as well as the total magnetization reduced with respect to its saturation value ($m/m_{s}$) are plotted in Fig.~\ref{MAG}(a)-(d) against the external magnetic field for one selected value of the RZFS parameter $E/\left|J\right|=0.5$ and four different values of the AZFS parameter. The value $D/\left|J\right|=-0.6$ is chosen so as to achieve the quantum ferrimagnetic (paramagnetic) phase $\left| FP \right\rangle$ ($\left| PP \right\rangle$) in the ground state at low (high) enough magnetic fields. In this particular case, the total magnetization at first rises steadily as the longitudinal magnetic field strengthens until it reaches 
a certain critical field at which one observes the magnetization jump due to the abrupt spin reversal of all the spin-$1/2$ atoms into the external-field direction. The initial continuous increase of the total magnetization as well as the one observed after the critical field can evidently be attributed 
to a gradual increase of the occurrence probability of the majority spin state $\left|S_{k}^z=1\right\rangle$, which appears in $\left| FP \right\rangle$ or 
$\left| PP \right\rangle$ at the expense of the occurrence probability of the minority spin state $\left|S_{k}^z=-1\right\rangle$ owing to the magnetic field strengthening. 
It is noteworthy that the strict magnetization jump serving in evidence of the field-induced transition from $\left| FP \right\rangle$ to $\left| PP \right\rangle$ takes place just at zero temperature, 
however, the rather steep (but still continuous) increase of the magnetization in a close neighborhood of the critical field reflects this zero-temperature transition at sufficiently low (but non-zero) temperatures as well. Furthermore, the second selected value of the AZFS parameter $D/\left|J\right|=-1.4$ is chosen strong 
enough in order to cause an appearance of the disordered (paramagnetic) phase $\left| DP \right\rangle$ ($\left| PP \right\rangle$) in the ground state at sufficiently low (high) magnetic fields. In this special case, the total magnetization exhibits at low magnetic fields the intermediate magnetization plateau at one third of the saturation magnetization, which only comes from a perfect alignment of the spin-1/2 atoms into the external-field direction (the spin-1 atoms do not contribute to the total magnetization at all as they reside the non-magnetic state $\left|S_{k}^z=0\right\rangle$ within $\left| DP \right\rangle$). At a certain critical field, the quantum spin chain undergoes the field-induced transition from $\left| DP \right\rangle$ to $\left| PP \right\rangle$, which is accompanied with the abrupt change in the spin state of the spin-$1$ atoms from the non-magnetic state $\left|S_{k}^z=0\right\rangle$ to the entangled pair of states $\left|S_{k}^z= \pm 1\right\rangle$. In the consequence of that, the total magnetization rises steadily with the magnetic field after the relevant critical field owing to a gradual change in the occurrence probabilities of the majority and minority spin states $\left|S_{k}^z=1\right\rangle$ and $\left|S_{k}^z=-1\right\rangle$, which appears in  $\left| PP \right\rangle$ under the influence of the longitudinal magnetic field. 
 
It should be nevertheless mentioned that the most interesting magnetization curves can be detected 
if a relative strength of the AZFS parameter is selected from the relatively narrow interval $D/\left|J\right|\in(-1.115,-1.040)$ when assuming $E/\left|J\right|=0.5$. Figs.~\ref{MAG}(c),(d) provide two illustrative examples of the low-temperature magnetization curves with the AZFS parameter taken from this range, which ensures occurrence of two sequential field-induced transitions from $\left| FP \right\rangle$ to $\left| DP \right\rangle$ and subsequently from $\left| DP \right\rangle$ to $\left| PP \right\rangle$ upon the magnetic field strengthening. Actually, it can be clearly seen from Figs.~\ref{MAG}(c),(d) that the total magnetization at first continuously increases within $\left| FP \right\rangle$ with increasing the magnetic field until it reaches the lower critical field, where the magnetization jumps towards the intermediate plateau with the constant value of the total magnetization that is stable for mediate values of the external magnetic field. The plateau region, which corresponds to the appearance of $\left| DP \right\rangle$, then ends up at the upper critical field reflecting another field-induced transition from $\left| DP \right\rangle$ to $\left| PP \right\rangle$. Accordingly, the total magnetization jumps from one third of the saturation value to a rather high value relatively close to the saturation magnetization and then, it rises steadily with increasing the longitudinal magnetic field within $\left| PP \right\rangle$ until it reaches the saturation magnetization in the limit of infinitely strong magnetic field 
$H/\left|J\right| \to \infty$.  

\begin{figure} 
\vspace{-0.8 cm}
\begin{center}
\includegraphics[width = 0.49\textwidth]{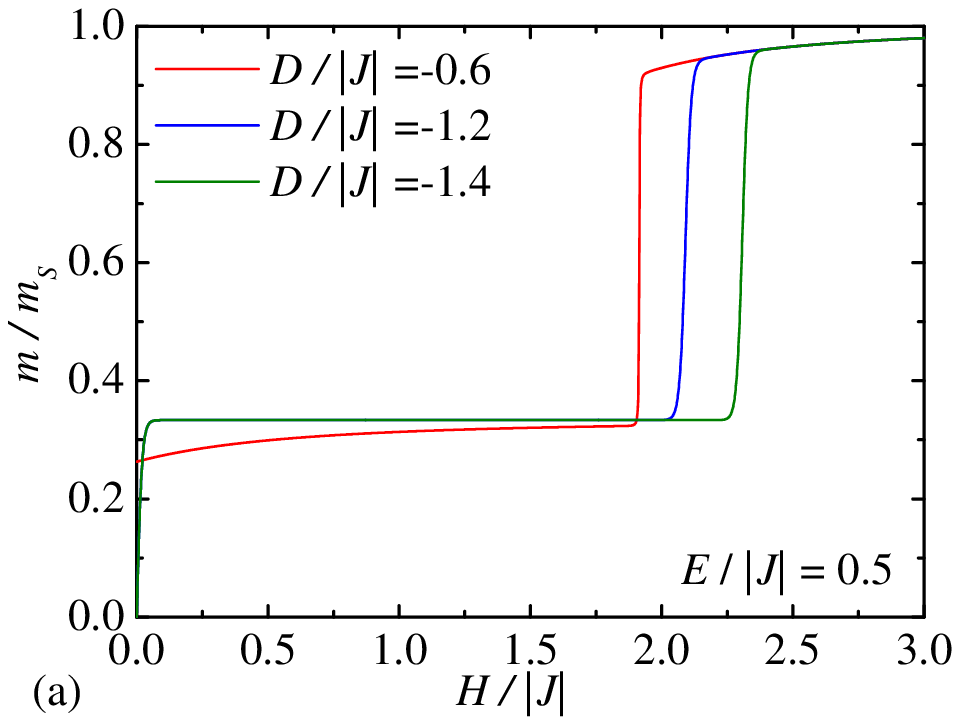}
\includegraphics[width = 0.49\textwidth]{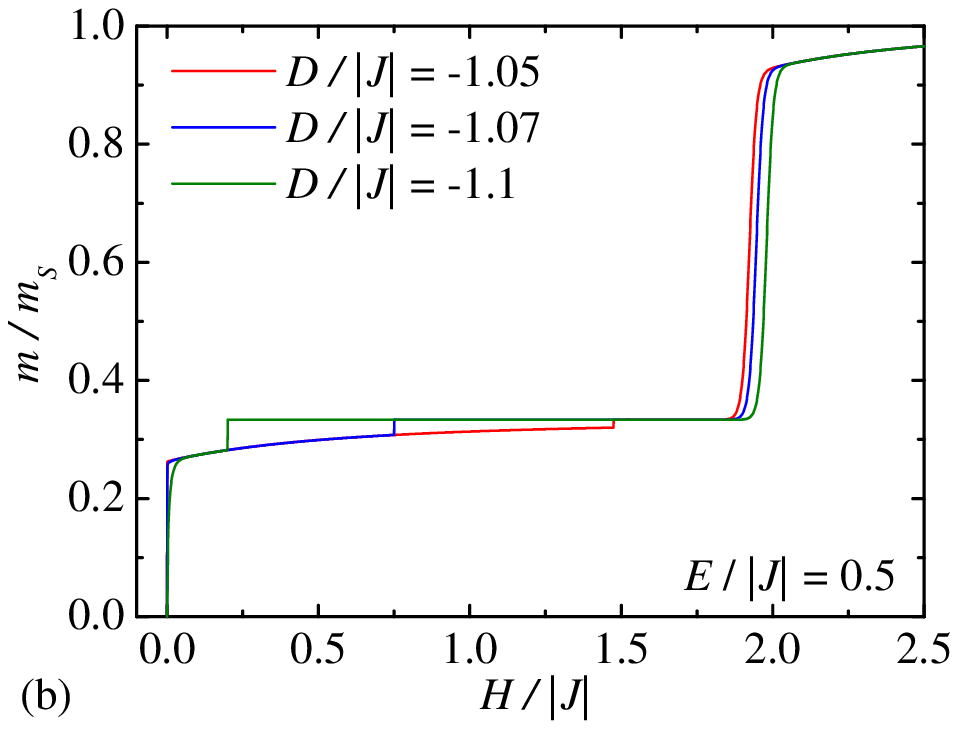}
\vspace{-0.6 cm}
\caption{\small The total magnetization reduced with respect to its saturation value as a function of the magnetic field at the sufficiently low temperature $k_{B}T/\left|J\right|=0.01$, one specific value of the RZFS parameter $E/\left|J\right|=0.5$ and several values of the AZFS parameter: (a) $D/\left|J\right| = -0.6, -1.2, -1.4$; (b) $D/\left|J\right| = -1.05, -1.07, -1.1$.}
\label{MAG2}
\end{center}
\end{figure}

For the sake of comparison, the total magnetization reduced with respect to its saturation value is plotted against the external magnetic field in Figs.~\ref{MAG2}(a),(b) for one specific value of the RZFS parameter $E/\left|J\right|=0.5$ and several values of the AZFS parameter at low enough temperature $k_{B}T/\left|J\right|=0.01$. These figures clearly demonstrate a possible diversity in the magnetization process, which might include the regions with the continuous change in the total magnetization, the intermediate magnetization plateau, as well as, one or two magnetization jumps accompanying the field-induced transitions between different phases. Finally, it should be also noted here that the temperature rise gradually smoothens the magnetization behavior close to and at the intermediate magnetization plateau, which cannot be consequently seen in the relevant magnetization curves at sufficiently high temperatures.

\subsection{Specific heat}

\begin{figure} 
\vspace{-0.8 cm}
\begin{center}
\includegraphics[width = 0.49\textwidth]{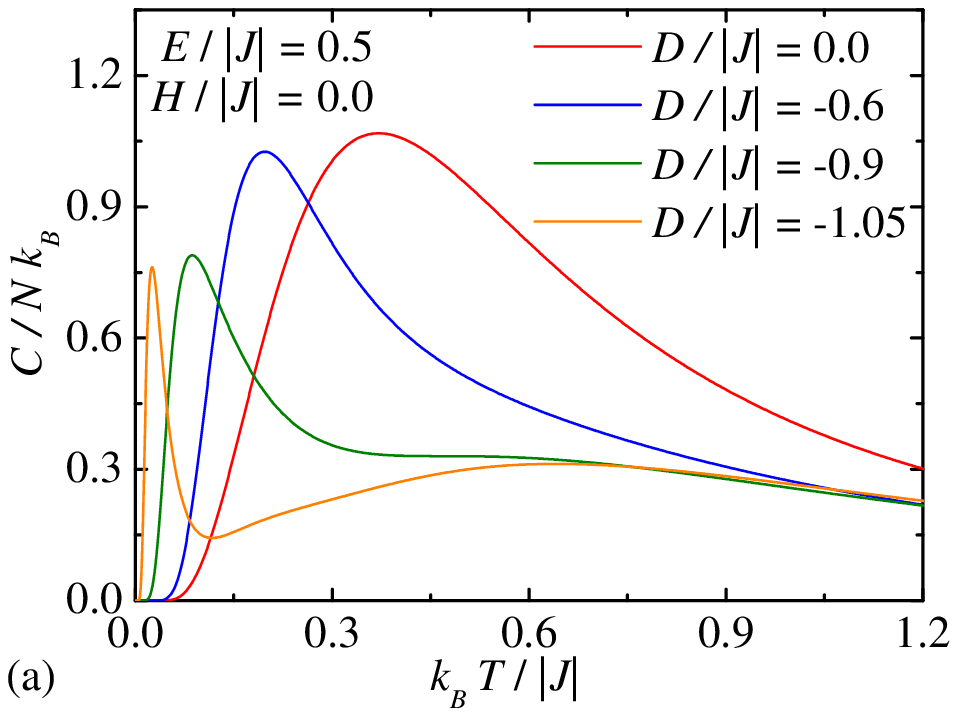}
\includegraphics[width = 0.49\textwidth]{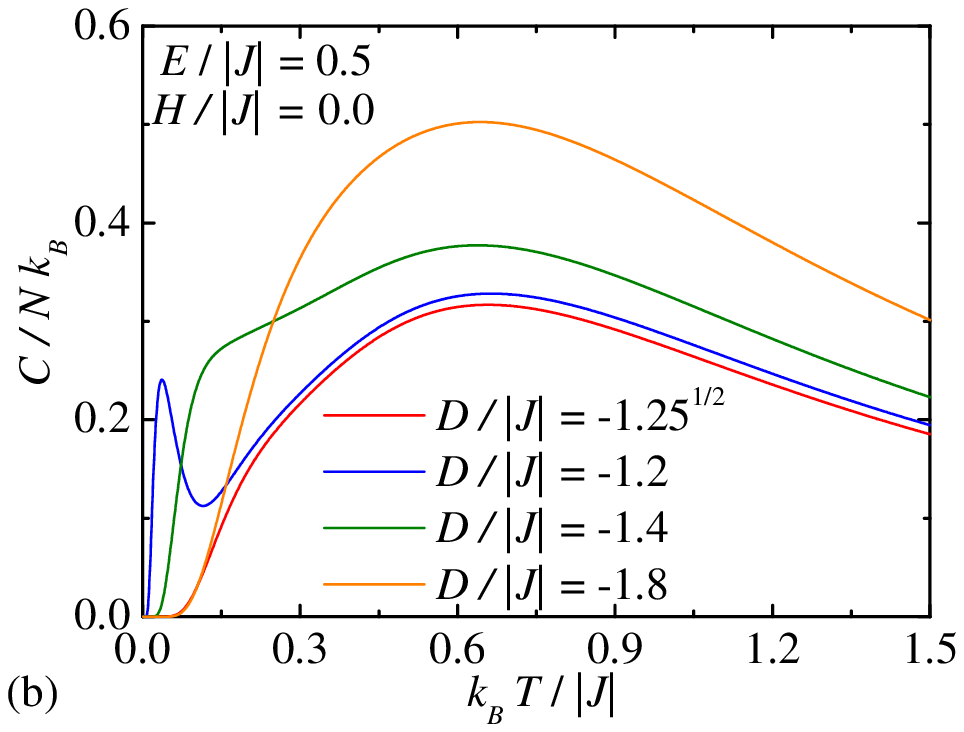}
\includegraphics[width = 0.49\textwidth]{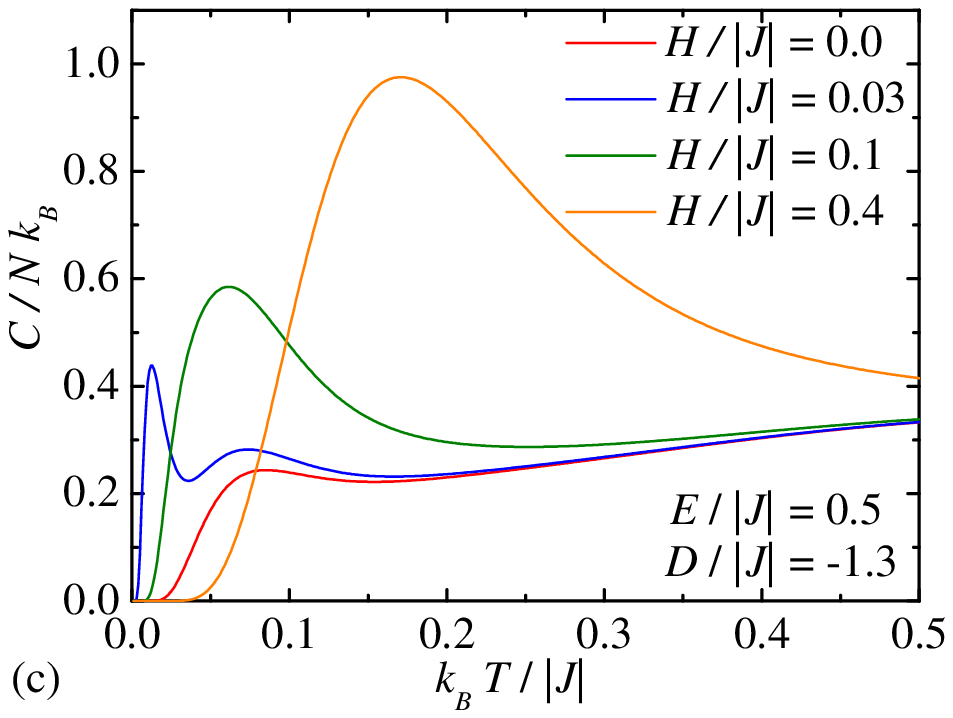}
\includegraphics[width = 0.49\textwidth]{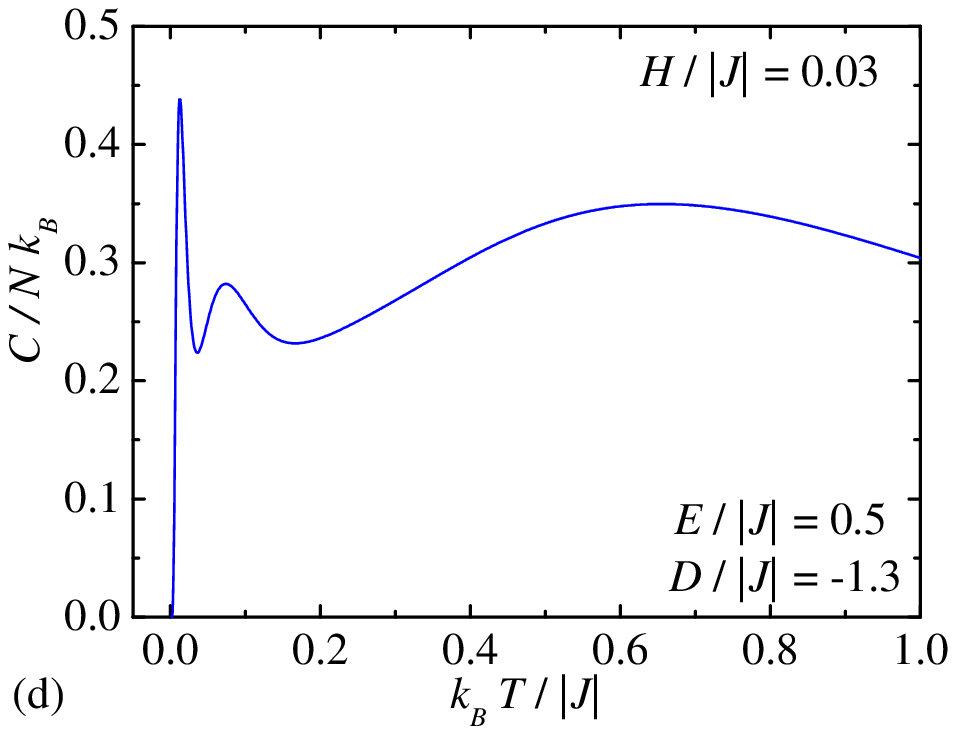}
\vspace{-0.6 cm}
\caption{\small Temperature changes of the specific heat for one selected value of the RZFS parameter  $E/\left|J\right|=0.5$ and several values of the AZFS parameter. Figs. \ref{SH}(a)-(b) show thermal variations of the specific heat in the zero-field case $H/\left|J\right|=0.0$, whereas the special case $D/\left|J\right|= -\sqrt{1.25}$ corresponds to the phase boundary between $\left| FP \right\rangle$ and $\left| DP \right\rangle$. Figs. \ref{SH}(c)-(d) illustrate thermal variations of the specific 
heat in the non-zero magnetic field when assuming $D/\left|J\right|=-1.3$.}
\label{SH}
\end{center}
\end{figure}

Another quantity, which is important for overall understanding of thermodynamics, is the magnetic contribution to the specific heat. Temperature variations of the zero-field specific heat are shown in Figs.~\ref{SH}(a),(b) for the RZFS parameter $E/\left|J\right|=0.5$ and several values of the AZFS parameter. As one can see, the temperature dependence of the specific heat always exhibits at least one round Schottky-type maximum regardless of whether $\left| FP \right\rangle$ or $\left| DP \right\rangle$ constitutes the ground state.
The specific heat has a single rounded maximum when considering either positive or small negative values of the AZFS parameter warranting the appearance of $\left| FP \right\rangle$ in the ground state such as $D/\left|J\right| = 0.0$ and $-0.6$. If the AZFS parameter is selected close enough to the phase boundary between $\left| FP \right\rangle$ and $\left| DP \right\rangle$, then, the additional second peak can be observed at low temperatures no matter whether $\left| FP \right\rangle$ or $\left| DP \right\rangle$ constitutes the ground state (see the specific heat curves for $D/\left|J\right| = -1.05$ and $-1.2$). However, the pronounced double-peak dependence of the specific heat gradually changes into the dependence with a single more or less symmetric maximum when considering more negative values of the AZFS parameter far from the phase boundary between $\left| FP \right\rangle$ and $\left| DP \right\rangle$ (see the thermal dependences for $D/\left|J\right| = -1.4$ and $-1.8$). These observations would suggest that the low-temperature peak in the double-peak dependence of the zero-field specific heat arises from thermal excitations between the ground-state spin configuration and the relevant excited state with only slightly higher energy.

The situation becomes even more intriguing when assuming the AZFS parameter close enough to the phase boundary between $\left| FP \right\rangle$ and $\left| DP \right\rangle$ and applying the non-zero magnetic field. It is quite apparent from Fig.~\ref{SH}(c) that the thermal dependence of specific heat with a single more or less symmetric maximum can be recovered upon strengthening the external magnetic field (see the curve for $H/\left|J\right|=0.5$). However, one may also detect the very unusual triple-peak dependence of the heat capacity when applying the sufficiently small but non-zero longitudinal magnetic field as depicted in Fig.~\ref{SH}(d). It is noteworthy that the third peak is located in the low-temperature area and the emergence of this additional third peak can be explained as the Zeeman's splitting of energy levels. In agreement with this argument, the third peak shifts linearly towards higher temperatures with increasing the strength of the magnetic field until it completely merges 
with the second low-temperature maximum. Therefore, the triple-peak dependence of the heat capacity 
may be observed in a relatively weak magnetic fields only.

\subsection{Longitudinal susceptibility}

\begin{figure} 
\vspace{-0.8 cm}
\begin{center}
\includegraphics[width = 0.49\textwidth]{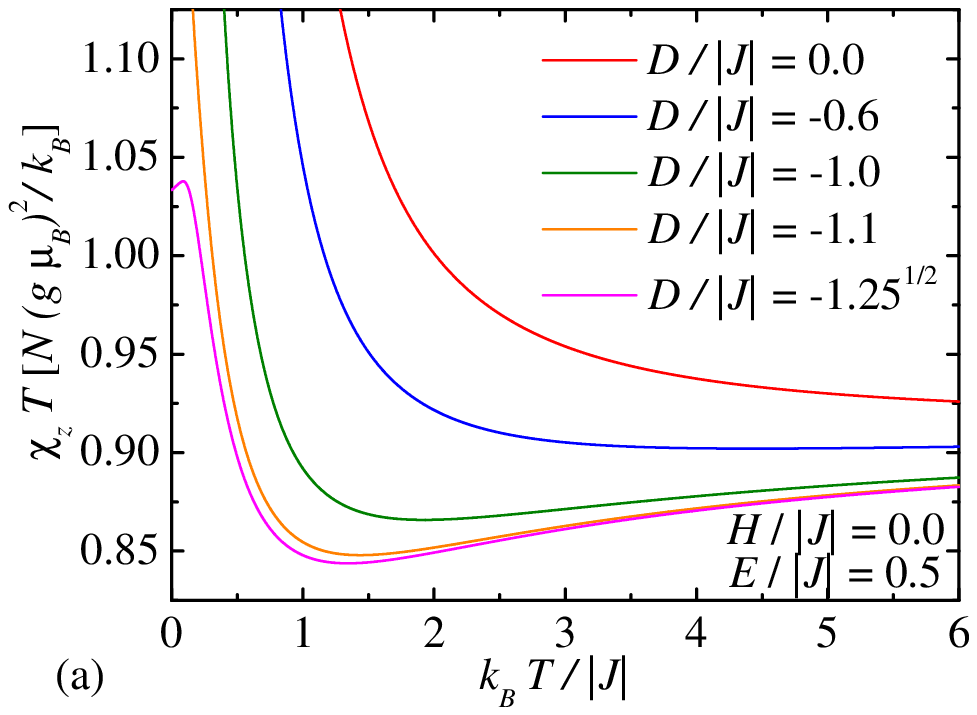}
\includegraphics[width = 0.49\textwidth]{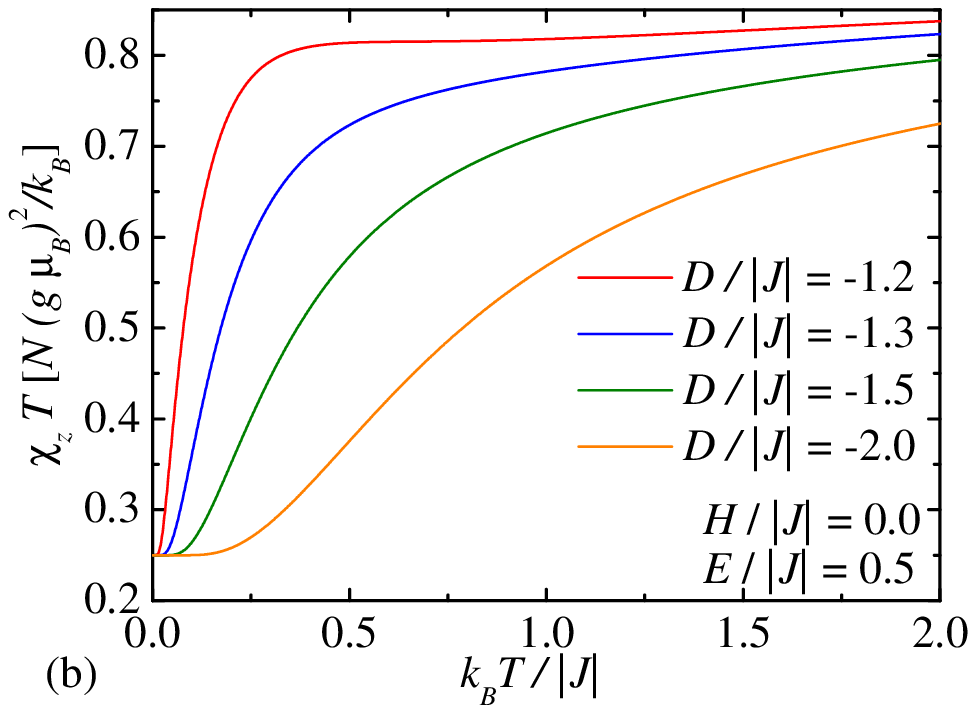}
\vspace{-0.6 cm}
\caption{\small The temperature dependence of the initial longitudinal susceptibility times temperature ($\chi_z T$) product for one selected value of the RZFS parameter $E/\left|J\right|=0.5$ and several values of the AZFS parameter. The particular case $D/\left|J\right|= -\sqrt{1.25}$ corresponds to the phase boundary between $\left| FP \right\rangle$ and $\left| DP \right\rangle$.}
\label{SUS}
\end{center}
\end{figure}

Next, let us examine the thermal dependence of the initial longitudinal susceptibility times temperature product $\chi_z T$ as shown in Fig.~\ref{SUS}. If the AZFS parameter is selected so as to achieve $\left| FP \right\rangle$ in the ground state, then, $\chi_z T$ product generally exhibits an exponential divergence when approaching zero temperature as depicted in Fig.~\ref{SUS}(a). It can be seen from this figure that $\chi_z T$ monotonically decreases with increasing temperature for positive or small negative values of the AZFS parameter. In addition, one may also detect a more striking non-monotonous temperature dependence of $\chi_z T$ with a rather flat minimum when the AZFS parameter 
is chosen sufficiently close but slightly above the ground-state boundary between $\left| FP \right\rangle$ and $\left| DP \right\rangle$ [see for instance the curve $D/\left|J\right|= -1.1$ 
in Fig.~\ref{SUS}(a)]. The emergence of the round minimum in the temperature dependence of $\chi_z T$ product is a typical feature of the quantum ferrimagnets \cite{yama98}, because the monotonous 
decrease (increase) of $\chi_z T$ product with increasing temperature indicates predominantly ferromagnetic (antiferromagnetic) character of excitations. The position of the round minimum can be thus regarded as a crossover point from the region with predominant ferromagnetic towards the region with predominant antiferromagnetic excitations. If the AZFS parameter is strong enough to stabilize $\left| DP \right\rangle$ in the ground state [see Fig.~\ref{SUS}(b)], then, $\chi_z T$ product   increases monotonically with increasing temperature from its minimum initial value $1/4$. This 
constant value evidently comes from the response of the spin-$1/2$ atoms with respect to the infinitesimal change of the magnetic field, because the spin-$1$ atoms occupy within the $\left| DP \right\rangle$ the non-magnetic spin state $\left|S_k^z=0\right\rangle$. It is also noteworthy that $\chi_z T$ product converges to the constant value of $11/12$ in the limit of high temperatures $T\to\infty$ regardless of whether $\left| FP \right\rangle$ or $\left| DP \right\rangle$ is 
being the ground state. This value agrees with the Curie law for the mixed spin-(1/2,1) system, 
which consist of an equal number of independent spins-$1/2$ and spin-$1$ atoms.

\subsection{Transverse susceptibility}

Further, let us proceed to a discussion of temperature dependences of the initial transverse susceptibility times temperature product. It is worthwhile to recall that the RZFS parameter effectively introduces a magnetic anisotropy into both transverse ($x$ and $y$) spatial components of the mixed-spin Ising chain refined by the AZFS and RZFS parameters and consequently, thermal variations of both spatial components of the initial transverse susceptibility ($\chi_x$ and $\chi_y$) might significantly differ one from another. 
Note furthermore that the positive value of the RZFS parameter corresponds to the situation when the $x$-axis is easier magnetization axis than the $y$-axis within the $xy$-plane. 

\begin{figure} 
\vspace{-0.8 cm}
\begin{center}
\includegraphics[width = 0.49\textwidth]{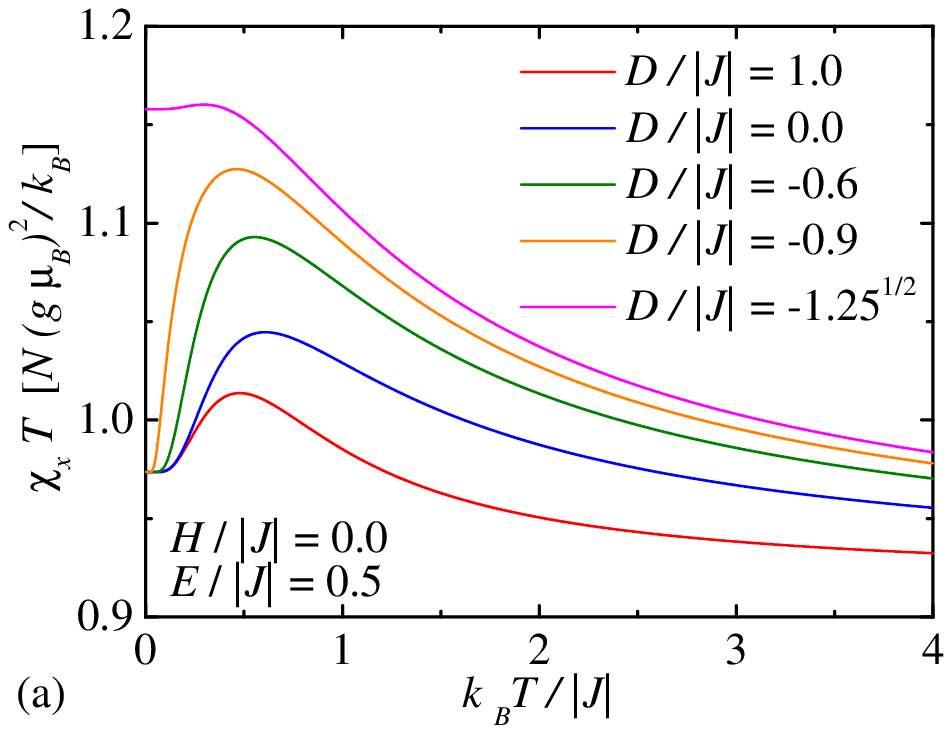}
\includegraphics[width = 0.49\textwidth]{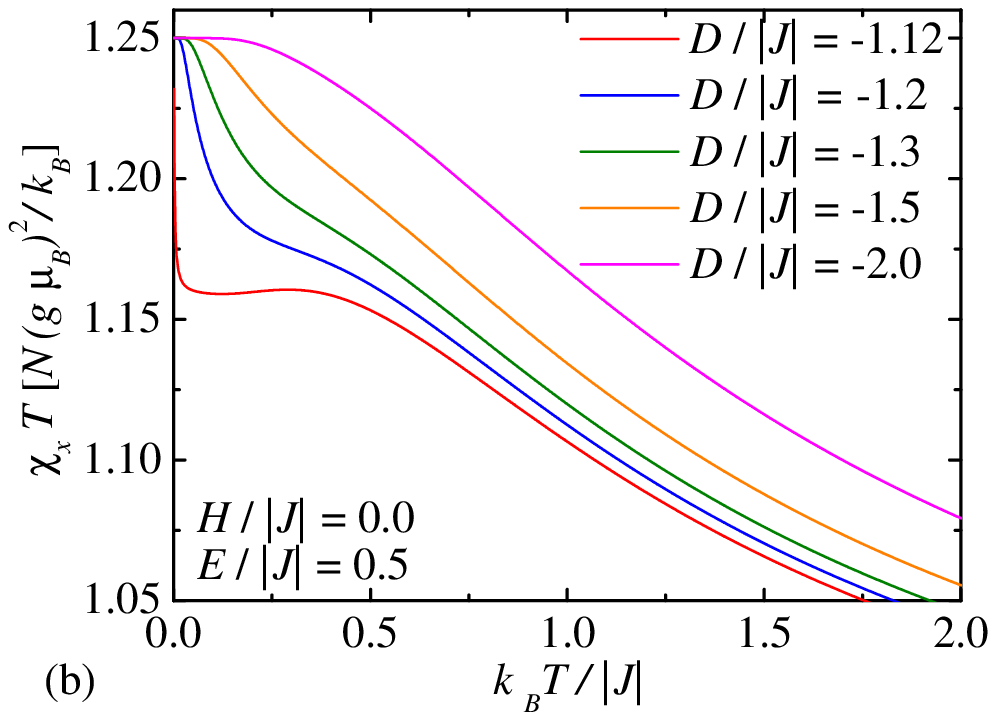}
\includegraphics[width = 0.49\textwidth]{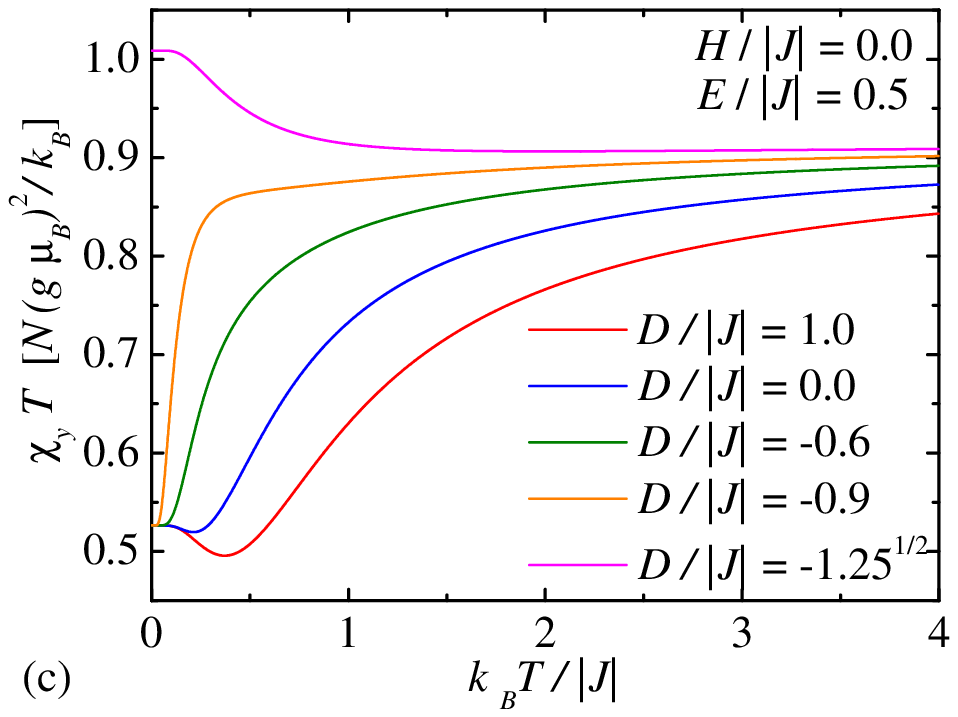}
\includegraphics[width = 0.49\textwidth]{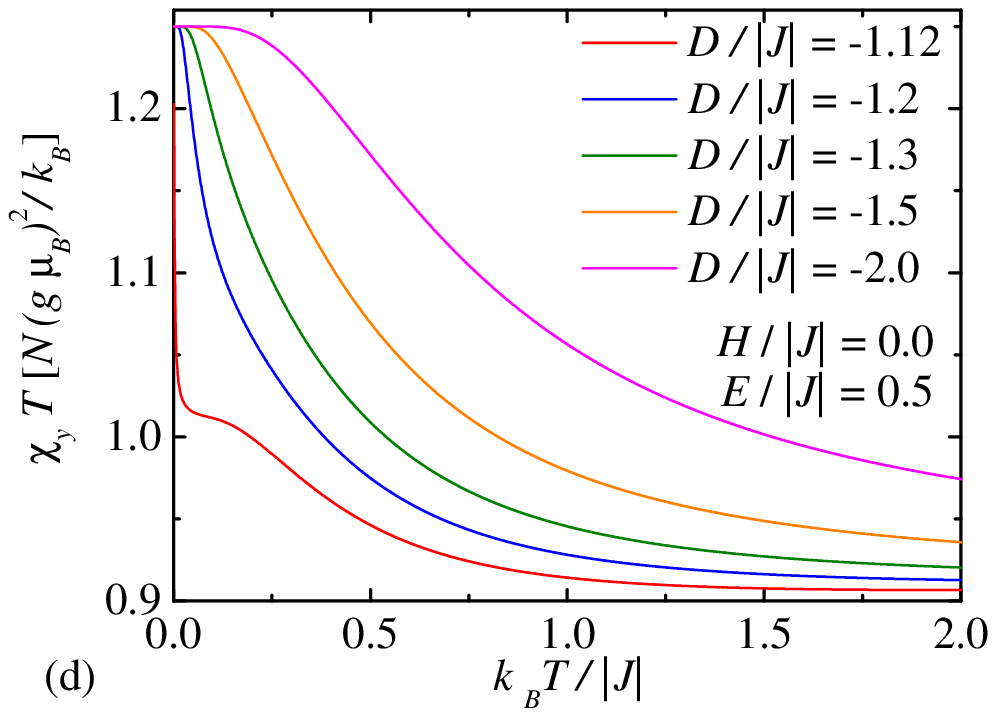}
\vspace{-0.6 cm}
\caption{\small Thermal variations of the initial transverse susceptibility times temperature product for the zero longitudinal magnetic field $H/|J|=0.0$, one particular value of the RZFS parameter $E/|J|=0.5$ and several values of the AZFS parameter. The upper (lower) panel shows the transverse susceptibility for the $x$-axis ($y$-axis), the left (right) panel illustrates the dependences 
for the ground state $\left| FP \right\rangle$ ($\left| DP \right\rangle$).}
\label{TS}
\end{center}
\end{figure}

First, let us compare thermal dependences of both initial transverse susceptibilities by considering the ground-state spin arrangement inherent to $\left| FP \right\rangle$ (the left panel in Fig.~\ref{TS}). 
It is noteworthy that both spatial components of the initial transverse susceptibility exhibit very different temperature variations of the susceptibility times temperature product in this particular case. $\chi_y T$ product initially decreases to its local minimum and then gradually increases with increasing temperature for positive or small negative values of the AZFS parameter (e.g. $D/|J|=1.0$ and $0.0$), while it monotonously increases upon temperature increase for more negative values of the AZFS parameter (e.g. $D/|J|=-0.6$ and $-0.9$). On the other hand, $\chi_x T$ product always exhibits a non-monotonous temperature dependence with a single rounded maximum. Another interesting observation is that $\chi_x T$ is in general greater than the corresponding value of $\chi_y T$ irrespective of temperature. According to Eq.~(\ref{24}), the greater (smaller) value of $\chi_x T$ ($\chi_y T$) relates to the greater (smaller) value of the $x(y)$-component of the quadrupolar moment $q_B^x$ ($q_B^y$). The zero-temperature limits of the transverse susceptibility times temperature product are actually consistent with the asymptotic values of the quadrupolar moment obtained for $\left| FP \right\rangle$ from Eq.~(\ref{19}) in the zero-temperature limit $T \to 0$
\begin{eqnarray}
q_B^x = \frac{1}{2} \left(1 + \frac{E}{\sqrt{J^2 + E^2}}\right), \quad 
q_B^y = \frac{1}{2} \left(1 - \frac{E}{\sqrt{J^2 + E^2}}\right). 
\label{qm}
\end{eqnarray}
This result is taken to mean that the $x$-axis is indeed easier magnetization axis than the $y$-axis in $\left| FP \right\rangle$. It is worthy to notice, moreover, that the zero-temperature value of transverse spatial component of the quadrupolar moment (\ref{qm}) has been confused with the zero-temperature value of the transverse susceptibility in the recent work of Wu \textit{et al}. \cite{hain08}.

It should be also mentioned that the difference in thermal dependences of both initial transverse susceptibilities becomes much less pronounced when considering the mixed spin-(1/2,1) Ising chain in the another possible ground state $\left| DP \right\rangle$ (the right panel in Fig.~\ref{TS}). The spin-1 atoms occupy within $\left| DP \right\rangle$ the non-magnetic spin state $\left|S_{k}^z=0\right\rangle$, which leads in turn to the equality of both transverse spatial components of the quadrupolar moment 
simultaneously achieving their maximum value $q_B^x = q_B^y = 1.0$ in the ground state. Owing to this fact, both spatial components of the initial transverse susceptibility times temperature product start from its maximum value $5/4$ and they must generally decrease with increasing temperature due to the temperature-induced lowering of the quadrupolar moment. The only non-monotonous temperature dependence 
of the initial transverse susceptibility times temperature product can be found in $\chi_x T$ vs. $T$ dependence if one considers the AZFS parameter close enough to the phase boundary between $\left| DP \right\rangle$ and $\left| FP \right\rangle$. In this special case, $\chi_x T$ initially decreases 
rather rapidly to its local minimum, then it rises steadily to a shallow local maximum and finally, it repeatedly exhibits a gradual decline with increasing temperature.

\subsection{Adiabatic demagnetization}

\begin{figure} 
\vspace{-0.8 cm}
\begin{center}
\includegraphics[width = 0.49\textwidth]{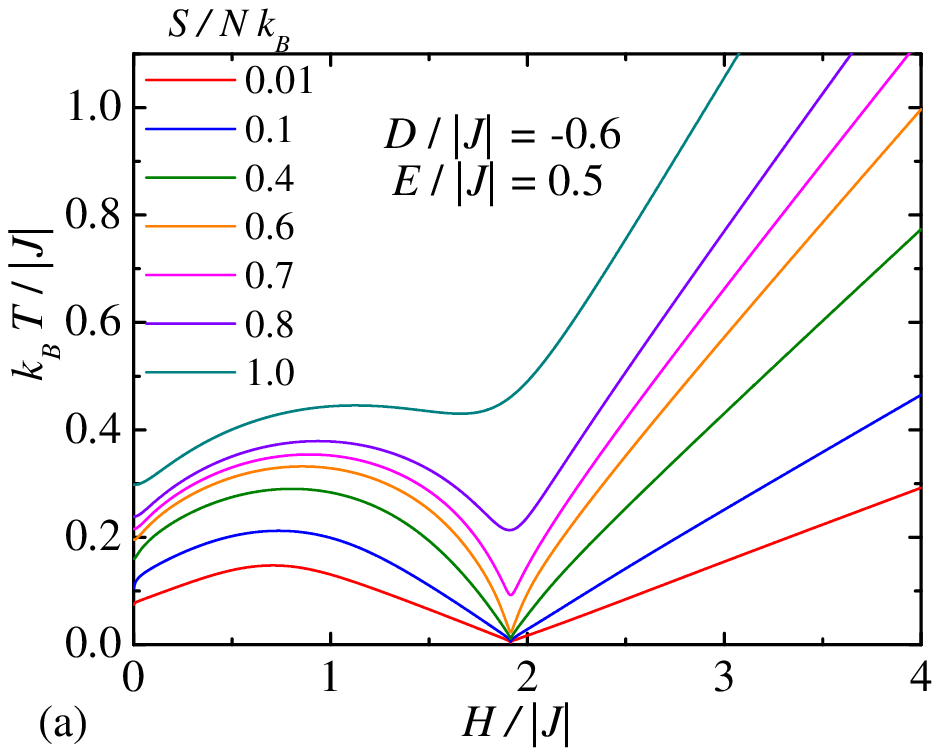}
\includegraphics[width = 0.49\textwidth]{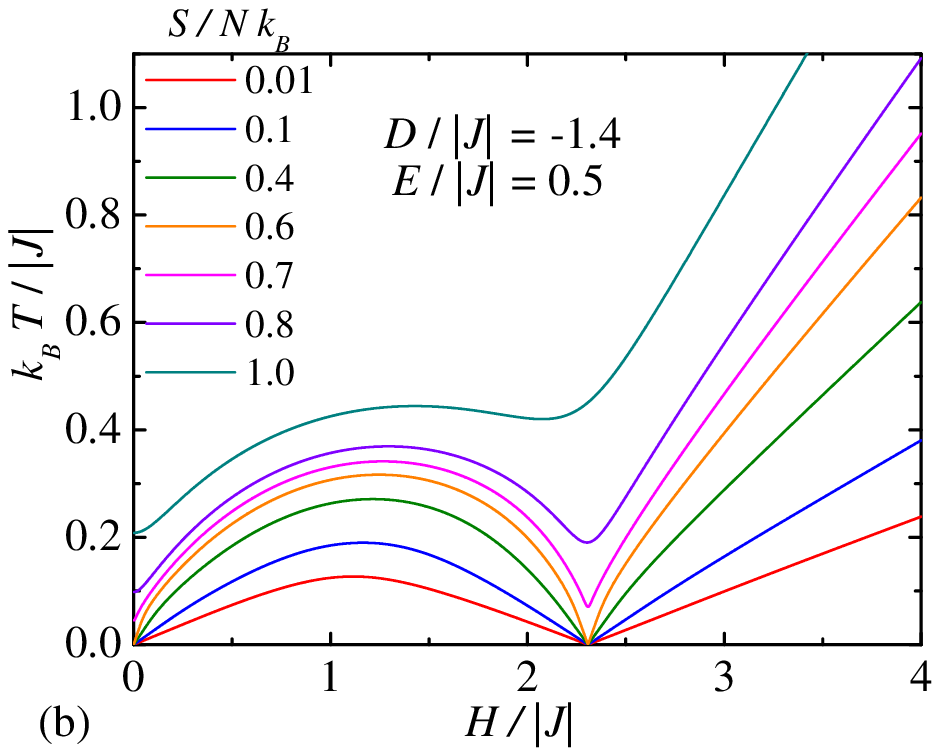}
\includegraphics[width = 0.49\textwidth]{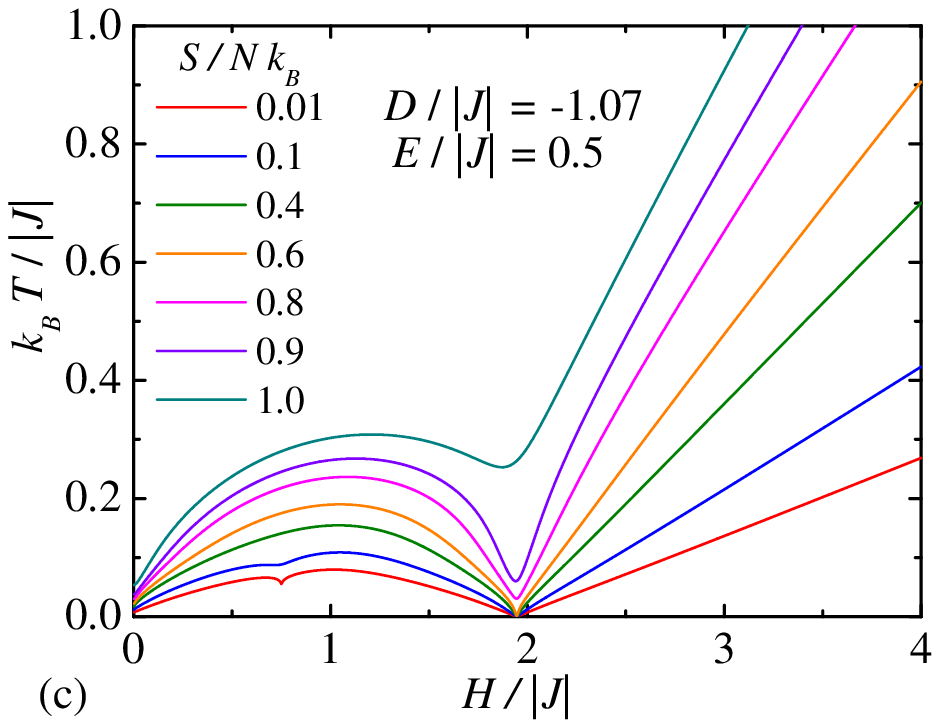}
\includegraphics[width = 0.49\textwidth]{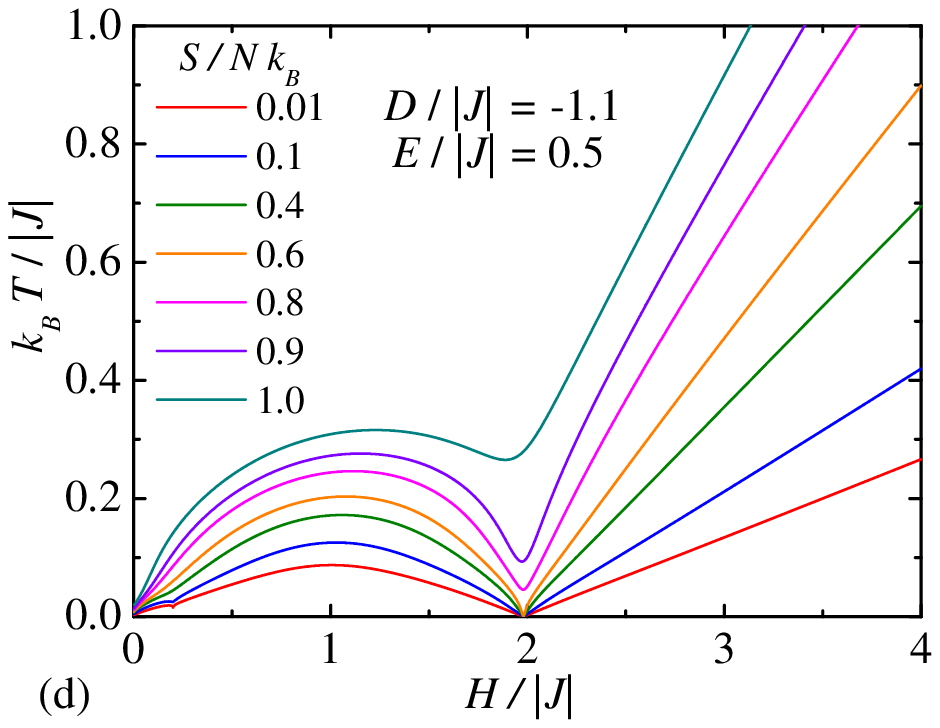}
\vspace{-0.6 cm}
\caption{\small Adiabatic changes of the temperature with the magnetic field for one selected value of the RZFS parameter $E/\left|J\right|=0.5$ and four different values of the AZFS parameter, which correspond to the magnetization curves depicted in Fig.~\ref{MAG}.}
\label{EN}
\end{center}
\end{figure}

\begin{figure} 
\vspace{-0.8 cm}
\begin{center}
\includegraphics[width = 0.49\textwidth]{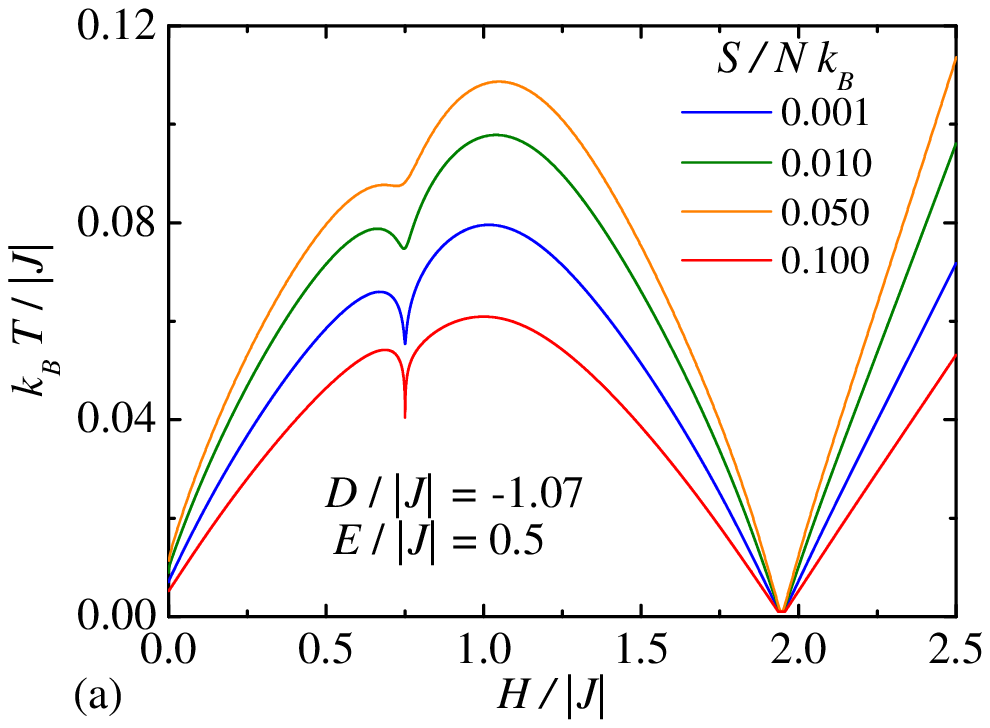}
\includegraphics[width = 0.49\textwidth]{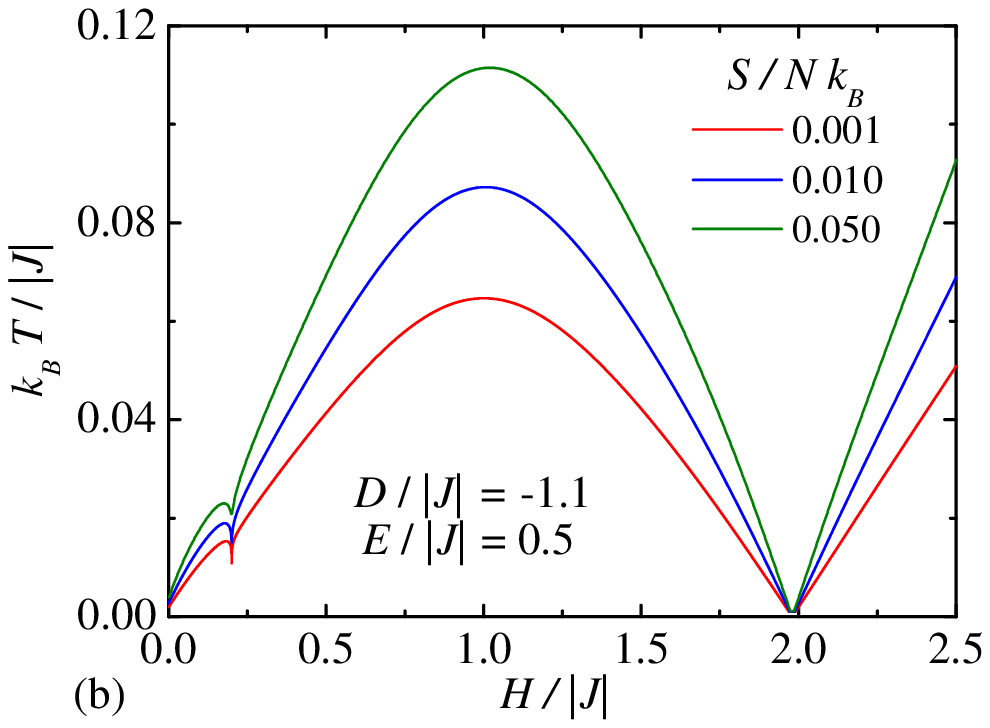}
\vspace{-0.6 cm}
\caption{\small Adiabatic changes of the temperature with the external magnetic field for one selected value of the RZFS parameter $E/\left|J\right|=0.5$ at relatively low values of the entropy. The small dip in the temperature versus the magnetic field dependence corresponds to the field-induced transition between $\left| FP \right\rangle$ and $\left| DP \right\rangle$.}
\label{ME}
\end{center}
\end{figure}

Last but not least, let us examine the adiabatic demagnetization in connection with a possibility of observing the enhanced magnetocaloric effect that might be of technological relevance for magnetic refrigeration \cite{zhit03,zhit04,hone05,derz05,cano09}. To gain an insight into the adiabatic demagnetization, Fig.~\ref{EN} displays the temperature as a function of the external magnetic field under the adiabatic condition (i.e. the constant entropy) for one specific value of the RZFS parameter $E/\left|J\right|=0.5$ and four different values of the AZFS anisotropy. Note furthermore that the displayed dependences correspond to the low-temperature magnetization curves depicted in Fig.~\ref{MAG}. 
The observed temperature changes can easily be understood from a comparison of Figs.~\ref{FD} and \ref{EN}. If the AZFS parameter is so selected that $\left| FP \right\rangle$ constitutes the zero-field ground state [Fig. \ref{EN}(a)], then, the spin system finally approaches a relatively 
low but non-zero temperature in the limit of vanishing external magnetic field. Contrary to this, 
the temperature tends towards the absolute zero when a sufficiently strong (negative) value of 
the AZFS parameter energetically stabilizes the disordered ground state $\left| DP \right\rangle$ 
in the zero-field limit $H/|J| \to 0.0$ [Figs.~\ref{EN}(b)-(d)]. From this point of view, the mixed spin-(1/2,1) Ising chain becomes a more efficient refrigerant if the AZFS parameter drives the system 
into the disordered ground state $\left| DP \right\rangle$. It is also worthy to notice that
the character of magnetocaloric effect basically changes at transition fields, where one 
generally observes a rather fast cooling (heating) when approaching the relevant transition field from above (below). Besides, it is also quite apparent from Fig.~\ref{EN} that the relatively fast and efficient cooling (heating) can be achieved only for the adiabatic processes with the constant entropy close enough to the value $S=N k_{\rm B} \ln 2$. Under this circumstance, the temperature falls rather quickly to zero if the external magnetic field is approaching either the transition field between $\left| FP \right\rangle$ and $\left| PP \right\rangle$, $\left| DP \right\rangle$ and $\left| PP \right\rangle$, or it tends towards zero. On the other hand, the temperature reaches just some non-zero (even if relatively small) value if the external magnetic field approaches the transition field between $\left| FP \right\rangle$ and $\left| DP \right\rangle$. 
It is noteworthy that the latter behavior cannot be clearly seen within the displayed scale from Fig.~\ref{EN} and thus, it is separately demonstrated in Fig.~\ref{ME}. It is quite evident from this figure that a relatively small dip (minimum) can be observed at the field-induced transition between $\left| FP \right\rangle$ and $\left| DP \right\rangle$, whereas this local minimum in the magnetic field dependence of the temperature already disappears upon a small increase of the entropy of 
the adiabatic process. 

\section{Concluding remarks}
\label{conclusion}

In this article, the mixed spin-(1/2,1) Ising chain with the AZFS and RZFS parameters in the applied longitudinal magnetic field has been exactly solved by combining the decoration-iteration transformation with the transfer-matrix method. Exact results for phase diagrams, free energy, magnetization, quadrupolar moment, entropy, specific heat, longitudinal and transverse initial susceptibilities have been derived and discussed in detail. It has been shown that the RZFS parameter basically modifies the magnetic behavior of the investigated mixed-spin chain, since it is responsible for the quantum entanglement between two magnetic states $| S_k^z = \pm 1 \rangle$ of the spin-1 atoms. 

Among the most interesting findings to emerge from the present study one could mention an extraordinary diverse magnetization process, which may include the regions with the continuous change in the total magnetization, the intermediate magnetization plateau at one third of the saturation magnetization, 
as well as, one or two magnetization jumps accompanying the field-induced transitions between different phases. Another interesting findings concern with an appearance of the round minimum in the temperature dependence of susceptibility times temperature data, the double-peak zero-field specific heat curves and the enhanced magnetocaloric effect in a vicinity of the field-induced phase transitions. It has been evidenced that the triple-peak temperature dependence of the specific heat may also be found 
when driving the system through the AZFS and RZFS parameters close enough to a phase boundary between $\left| FP \right\rangle$ and $\left| DP \right\rangle$ and applying sufficiently small but non-zero longitudinal magnetic field. 

\section{Appendix}

The list of functions, which enter into the exact analytical formulas (\ref{is}) and (\ref{19}) for 
the initial longitudinal susceptibility and the quadrupolar moment in the limit of the vanishing longitudinal magnetic field $H_A = H_B \to 0$: 
\begin{eqnarray}
U_{1} &=& 1 + 2 \exp(\beta D) \cosh \left( \beta \sqrt{J^2+E^2} \right), \nonumber \\
U_{2} &=& 1 + 2 \exp(\beta D) \cosh \left( \beta E \right), \nonumber \\ 
U_{3} &=& \frac{2J}{\sqrt{J^2+E^2}} \exp(\beta D) \sinh \left( \beta \sqrt{J^2+E^2} \right), 
\nonumber \\
U_{4} &=& \frac{2 E^2}{\beta (J^2+E^2)^{3/2}} \exp(\beta D) \sinh \left( \beta \sqrt{J^2+E^2} \right), 
\nonumber \\
U_{5} &=& \frac{2J^2}{J^2+E^2} \exp(\beta D) \cosh \left( \beta \sqrt{J^2+E^2} \right), 
\nonumber \\
U_{6} &=& \frac{2}{\beta E} \exp(\beta D) \sinh \left( \beta E \right),
\nonumber \\
U_{1}^{\pm} &=& 1 + \exp(\beta D) \cosh \left( \beta \sqrt{J^2+E^2} \right)
 \pm \frac{E \exp(\beta D)}{\sqrt{J^2 + E^2}}  \sinh \left( \beta \sqrt{J^2+E^2} \right), 
\nonumber \\
U_{2}^{\pm} &=& 1 + \exp(\beta D) \cosh \left( \beta E \right) 
\pm \exp(\beta D) \sinh \left( \beta E \right). \nonumber   
\label{29}
\end{eqnarray}

%***************************************************************
%                     Literature
%***************************************************************

\end{document}